\documentclass[%
 reprint,
 amsmath,amssymb,
 aps,
]{revtex4-2}

\usepackage{graphicx}
\usepackage{dcolumn}
\usepackage{bm}
\usepackage{amsfonts}
\usepackage{color}
\begin{document}

\preprint{APS/123-QED}

\title{Superfluidity in the 1D Bose-Hubbard Model}

\author{Thomas G. Kiely}
 \altaffiliation{tgk37@cornell.edu}
\author{Erich J. Mueller}
 \altaffiliation{em256@cornell.edu}
\affiliation{Laboratory of Atomic and Solid State Physics, Cornell University, Ithaca, NY 14853}

\date{\today}

\begin{abstract}
We study superfluidity in the 1D Bose-Hubbard model using a %infinite 
variational matrix product state technique. We determine the superfluid density as a function of the Hubbard parameters by calculating the energy cost of phase twists in the thermodynamic limit. As the system is critical, correlation functions decay as power laws and the entanglement entropy grows with the bond dimension of our variational state. We relate the resulting scaling laws to the superfluid density. We compare two different algorithms for optimizing the infinite matrix product state and develop a physical explanation why one of them (VUMPS) is more efficient than the other (iDMRG). Finally, we comment on finite-temperature superfluidity in one dimension and how our results can be realized in cold atom experiments.

% relate them to the superfluid density.  

% , and relate it to 

% No BEC
% Gapless modes
% Critical

% What we do:
% Phase Diagram
% Superfluid fraction <-> Correlation functions
% Entanglement entropy scaling
% Connections to experiments

% How do we do it:
% VUMPS
% Thermodynamic limit
% Comparison between iDMRG and VUMPS

% Superfluidity in 1D is not associated with off-diagonal long-range order.

\end{abstract}

\maketitle

\section{\label{sec:intro}Introduction}
Superfluidity is one of the most spectacular  examples of macroscopic quantum coherence. 
%Broadly speaking, superfluidity is a collective phenomenon wherein a system exhibits zero-viscosity flow.
It is a collective effect where some fraction of the fluid flows without dissipation.
%It is characterized by the superfluid fraction, which proportion of the system flows without dissipation.
%; and (ii) the critical velocity, which is the maximum speed of dissipationless flow.
%quantifies the extent to which one can excite the system before superfluidity is destroyed. 
%In general the critical velocity can be substituted for other measures of energetic stability, such as the healing length, which may depend on the system's spatial geometry.
%It arises due to Bose-Einstein condensation, and 
In a Galilean-invariant system at zero temperature, this superfluid fraction is either 0 or 1~\cite{leggettZeroTempSFDensity}. In a lattice system, however, the superfluid fraction can take on intermediate values. Here we use matrix product state techniques~\cite{SCHOLLWOCKreview} to compute the zero-temperature superfluid fraction of the 1D Bose-Hubbard model as a function of its parameters: the chemical potential $\mu$, which controls the number of particles; the on-site interaction strength $U$; and the tunneling matrix element $t$. We reproduce the iconic Mott lobes in the $(\mu/t,U/t)$ plane, showing insulating regions where the superfluid fraction vanishes and superfluid regions where it is finite.  We connect the superfluid fraction to a number of other properties of the 1D lattice Bose gas.

Superfluidity in one dimension is special. In dimensions $d\geq 3$, superfluidity is usually accompanied by Bose-Einstein condensation, where the off-diagonal elements of the single particle density matrix, $\langle \psi^\dagger(r) \psi(r^\prime)\rangle$, approach a constant for large spatial separations, $|r-r^\prime|\to\infty$. This is a form of long-range order, corresponding to the spontaneous breaking of a continuous $\rm{U}(1)$ symmetry.  Long-range order of this form is not permitted in one dimension~\cite{hohenbergLRO,merminWagner}, so there is no condensation; rather, the density matrix falls off as a power law, $\langle \psi^\dagger(r) \psi(r^\prime)\rangle\sim(r-r^\prime)^{-K/2}$. We implement two variational matrix product state algorithms 
%a variant of the VUMPS algorithm, introduced by Zauner-Stauber \textit{et al.}~\cite{vumps}, 
which allows us to explicitly calculate these correlation functions in the thermodynamic limit \cite{mcculloch2008infinite,whiteDMRG1,whiteDMRG2,vumps,vumps2,whiteOneSiteDMRG}. A remarkable feature of one-dimensional superfluidity is that the exponent $K$ can be related to the superfluid density~\cite{delawareQMC,affleckQMC}. We numerically show this correspondence.

Our technique gives us access to the entanglement spectrum, which characterizes the quantum correlations between different parts of the system.  The effective low energy theory describing the 1D Bose-Hubbard model has a conformal invariance which leads to a scaling behavior of this spectrum~\cite{haldaneHarmonicFluid}.  We demonstrate this scaling in our data.

The 1D Bose-Hubbard model is iconic and has been very widely studied.  It is described by a Hamiltonian
\begin{equation}
    \mathcal{H}_{BH}=\sum_j\bigg(-t(a^\dagger_ja_{j+1}+h.c.)-\mu n_j+\frac{U}{2}n_j(n_j-1)\bigg)
    \label{eq:Hbh}
\end{equation}
where $ a_j$ are annihilation operators for particles on site $j$ and $n_j=a^\dagger_ja_j$. Unlike its fermionic cousin, the Bose Hubbard model is not integrable due to the infinitely-large local Hilbert space. This has made it a popular target for strong-coupling expansions~\cite{freericksBH,strongCoupling} and numerical techniques, such as quantum Monte Carlo (QMC) algorithms~\cite{qmc1,qmc2,qmc3}, density matrix renormalization group (DMRG) methods~\cite{pai,monien,monienNN,scholwock,Urba,ejima1,reentrance,itebd1,treeTensorNetwork}, the variational cluster approximation~\cite{vca1,vcaDDMRG}, exact diagonalization~\cite{exactDiag}, and even machine learning methods~\cite{matLearn1,matLearn2,rbm}. These prior works have largely focused on mapping out the phase diagram and, in particular, identifying the BKT transition point at the Mott lobe tip \cite{BKT1,BKT2,BKT3}. Calculating the superfluid fraction in this model has traditionally been challenging. A popular approach in the QMC community is to exploit an identity between the superfluid density and the imaginary-time winding number~\cite{pollockCeperley}. 
%QMC algorithms typically provide numerically exact results for finite-sized systems at finite temperature. As we will discuss in Section~\ref{sec:definitions}, one must be careful in relating mesoscopic results to the thermodynamic limit in one dimension. DMRG methods work directly in the zero-temperature limit, thereby evading some of these subtleties. The superfluid density has been computed in finite-length systems by imposing a phase twist to systems with open boundary conditions~\cite{pai,sfDrag1}. One must be careful, however, to account for the influence of the boundary conditions.
%is procedure 
%can be somewhat dubious due to the lack of translational invariance. 
DMRG techniques have been used to calculate the superfluid density  in finite-length systems by imposing a phase twist to systems with open boundary conditions~\cite{pai,sfDrag1},
or periodic boundary conditions~\cite{treeTensorNetwork,sfDrag2}.
%More recent advances in DMRG techniques~\cite{pbcs} have lead to the simulation of truly periodic systems~\cite{treeTensorNetwork,sfDrag2}, in which case the superfluid density is readily extracted with a boundary condition twist. 
%these systems, one must determine the finite-size scaling $\rho_s(L)$ to extract the infinite-length asymptotics. 
Our numerical method has some advantages over these prior approaches:  We 
minimize the energy within the space of translationally-invariant matrix product states, directly giving us results in the zero temperature and thermodynamic limit. 
%To the best of our knowledge, in no prior work has the superfluid density of the 1D Bose-Hubbard model been computed directly in the thermodynamic limit.

In Sec.~\ref{sec:LL} we give some required background about Luttinger liquid theory, which describes the low energy physics of our system.  In
Sec.~\ref{sec:definitions} we more precisely define superfluid density, $\rho_s$, in a 1D system.
We note that this definition  
is itself the subject of some debate, 
and we explain the relevant issues, the consensus, and the operational definition which will be used in this paper.  Section~\ref{sec:methods} describes our numerical approach.  We introduce two algorithms, iDMRG \cite{whiteDMRG1,whiteDMRG2}, and VUMPS \cite{vumps}.  We describe how to use these techniques to calculate $\rho_s$ and the relevant correlation functions.
%find that the most efficient approach is a generalization of the VUMPS algorithm introduced by Zauner-Stauber \textit{et al.}~\cite{vumps}.  We compare this algorithm to a more conventional optimization approach, iDMRG~\cite{}.
Section~\ref{sec:results} gives the results of our calculations, including a comparison of the convergence properties of iDMRG and VUMPS. Section~\ref{sec:exp} discusses %contemporary 
techniques for measuring superfluid density in 1D systems.
We summarize in Sec.~\ref{sec:conclusions}.

% which we address in Section~\ref{sec:definitions}. The position we take in this paper is that superfluidity in one dimension exists and that existing definitions agree at zero temperature, which is the subject of our investigation.  In Sec....

\section{\label{sec:LL}Luttinger liquid theory}
Here we review the most pertinent results from Luttinger liquid theory, as these are essential for our analysis and discussion. Luttinger liquid theory encompasses the low-energy descriptions of a large variety of interacting 1D systems~\cite{giamarchi}. 
%Its ubiquity owes to the divergence of the density-density susceptibility in one dimension, which signals that interactions will destabilize the non-interacting fixed point. 
For a bosonic system, the low-energy Hamiltonian can be derived~\cite{haldaneHarmonicFluid} by expanding the boson field operators as
\begin{equation}
    \psi(x)=\sqrt{\rho_0-\frac{1}{\pi}\nabla\theta(x)}e^{i\phi(x)},
    \label{eq:harmonicField}
\end{equation}
where $\rho_0$ is the average number density and $\nabla\theta(x)$ and $\phi(x)$ are canonically-conjugate fields corresponding to long-wavelength density and phase fluctuations, respectively. In terms of these fields, the Luttinger liquid Hamiltonian is of the form
\begin{equation}
    \mathcal{H}_{LL}=\frac{\hbar}{2\pi}\int dx~\big(v_j(\nabla\phi)^2+v_n(\nabla\theta-\pi\rho_0)^2\big).
    \label{eq:hll}
\end{equation}
This Hamiltonian describes gapless, long-wavelength fluctuations in the density and phase fields with respective sound velocities $v_n$ and $v_j$. The velocity of phase fluctuations is $v_j=\frac{\hbar\pi\rho_s}{m}$ where $\rho_s$ is the zero-temperature superfluid density, or equivalently the Drude weight (see Sec.~\ref{sec:definitions}). In a Galilean-invariant system, $\rho_s=\rho_0$ so that $v_j$ is not renormalized by interactions
-- 
which is consistent with the aformentioned theorem that the superfluid fraction of a translationally invarient systems is either zero or unity
%is the 1D analog of Leggett's more general theorem
~\cite{leggettZeroTempSFDensity}. 
The velocity of density fluctuations is $v_n=1/\hbar\pi \kappa$ where $\kappa=\partial \rho_0/\partial\mu$ is the charge compressibility. It is common practice to re-parameterize Eq.~(\ref{eq:hll}) in terms of a single sound velocity, $u=\sqrt{v_jv_n}$, and the dimensionless ``Luttinger parameter," $K=\sqrt{v_n/v_j}$. Diagonalizing the Hamiltonian with a Bogoliubov transformation yields~\cite{haldaneHarmonicFluid}
\begin{equation}
    \mathcal{H}_{LL}=\hbar\bigg(\sum_{q\neq 0}\omega_qb^\dagger_qb_q+\bigg(\frac{\pi}{2L}\bigg)(v_jJ^2+v_n(N-N_0)^2)\bigg)
    \label{eq:diagHLL}
\end{equation}
where $\omega_q=u|q|$ for small $q$ and $b_q$ ($b^\dagger_q$) are the Bogoliubov annihilation (creation) operators. We can therefore see that excitations of the Luttinger liquid are sound modes that are a linear combination of density and phase fluctuations. The parameters $J$ and $N$ correspond to the total number of $\pi$-phase twists and total number of particles, respectively, over the length $L$ of the system. Periodic boundary conditions on the bosonic many-body wavefunction imply $J\in 2\mathbb{Z}$. The average number of particles is given by $N_0$.

The Luttinger liquid has a host of interesting properties. Despite being a bosonic theory, the lack of long-range order in one dimension prevents Bose-Einstein condensation. The propensity to order nonetheless leads to a power-law decay of the single-particle equal-time Green's function, $\langle a^\dagger_ia_{i+j}\rangle$%, where $a_i$ is the annihilation operator for the UV bosons (in this case described by the Bose-Hubbard Hamiltonian)
~\cite{giamarchi,haldaneHarmonicFluid}:
\begin{equation}
    \langle a^\dagger_ia_{i+j}\rangle\approx n_0(n_0 j)^{-K/2}.
    \label{eq:dMatLL}
\end{equation}
Here $n_0=\rho_0 d$ is the average number of particles per site, where $d$ is the lattice spacing.
Power-law behavior is also observed in a variety of other correlation functions, such as the density-density correlation function.
%: $d(x)=\langle n_in_{i+x}\rangle$, where $n_i=a^\dagger_ia_i$ is the number operator. These are summarized as
% \begin{eqnarray}
%     g(x)&\approx&\rho_0(\rho_0x)^{-K/2}\\
%     d(x)&\approx&\rho_0^2-\frac{2}{K}\frac{1}{(2\pi x)^2}+A(\rho_0x)^{-2/K}\cos(2\pi\rho_0x)
% \end{eqnarray}
%where $A$ is a constant that depends on short-ranged, non-Luttinger-liquid behavior, and we have omitted higher-order terms~\cite{haldaneHarmonicFluid}. 
The exponents depend on the Luttinger parameter, $K$, and in that sense they are ``tunable" functions of the number density and interaction strength. As the Luttinger parameter determines the long-distance behavior of the correlation functions, it's value also determines the propensity of the system to order in different ways. The single-component 1D Bose-Hubbard model hosts two phases: a Mott insulating phase and a superfluid (Luttinger liquid) phase.
%it also determines the relative propensity of the system to different kinds of ordering. The single-component 1D Bose-Hubbard model has only one kind of ordered phase, the Mott insulator (note this breaks a discrete symmetry and is thus permitted in 1D~\cite{hohenbergLRO,merminWagner}). 
In the superfluid phase $K<1$ while  at the SF-Mott transition $K\to 1$; the only exception is at the Mott lobe tip, where the SF-Mott transition is in the XY universality class~\cite{fisherBH} and the system undergoes a BKT transition~\cite{BKT1,BKT2,BKT3} in which $K\to 1/2$.

\section{\label{sec:definitions}Defining superfluid density}
One of our goals is to clearly articulate the subtleties arising in 1D superfluids. As prefaced in the introduction, superfluidity in one dimension is ``unconventional." 
Not only is
%insofar as 
Bose-Einstein condensation absent in these systems,
but the very definition of superfluid fraction has
ambiguities.
%does not occur. 
%While formal definitions of the superfluid fraction~\cite{sfCriterion,fisherHelicityModulus} do not require the existence of a background condensate, 
%and indeed are perfectly well-defined in the absence of condensation in two spatial dimensions, 
%Additionally, basic complications arise in defining the superfluid fraction in one dimension. 
We emphasize that this is not merely an issue of theoretical importance: as we show in Section~\ref{sec:neq}, this has led to a discrepancy between theory and experiment that necessitates a more nuanced understanding of 1D superfluidity.

It will be useful to have a concrete picture in mind. For the purpose of this section, we will imagine a 1D channel of length $L$, which forms a ring with radius $R=L/2\pi$. We will consider some artificial magnetic flux threading the ring, or equivalently a vector potential that points along the channel. The flux induces a current, and the current response defines the superfluid fraction. One can imagine equilibrium and non-equilibrium formulations of this thought experiment~\cite{leggettSF}. In the former, one inserts a small amount of flux and allows the system to come to its true ground state. If the flux is small enough, the resulting state will carry a finite current whose magnitude is proportional to the flux. This is known as the Hess-Fairbank effect~\cite{hessFairbank}.  Fundamentally it is a mesoscopic effect because the equilibrium current is a periodic function of the flux, and the relevant magnetic fields scale as $1/R$.
% Aharanov-Bohm phase exceeds $\pi/2$, the current carried by the superfluid 
% the flux needed to excite a vortex vanishes as $1/R$ when the length of the chain is increased. 
The non-equilibrium formulation involves first allowing the system to equilibrate in the presence of a large magnetic field.  One then turns off the magnetic field.  Assuming friction with the walls, a normal fluid rapidly come to rest.  A superfluid will not.  Typically one expects that the superfluid fraction measured via these two approaches will agree~\cite{leggettSF}. This is not the case in one dimension.

% Alternatively, one might imagine exciting the system with a large amount of flux. Letting the system equilibrate and then turning off the magnetic field, one would find that the normal component comes to rest while the superfluid component rotates for astronomically long timescales. This is known as metastable superflow. It is expected that both methods should agree in their definition of a superfluid fraction, at least in the thermodynamic limit~\cite{leggettSF}. We will see that this is not the case in one dimension. 
%As mentioned in the introduction, a superfluid is also defined by a critical velocity $v_c$ above which the excitations will lead to viscous flow~\cite{landau1}. In general, then, the induced angular velocity in the 1D chain (interpreted as motion around a 2D ring) must satisfy $\omega<v_c/R$. Rotation is quantized on a ring, however, so the angular velocity must also obey 
%In this section we will summarize recent work on this subject and define precisely how the quantity we refer to as the superfluid density relates to other definitions.

\subsection{\label{sec:drude}Drude weight, superfluid density, and helicity moduli}
One of the subtleties we need to contend with is the formal similarities between a superfluid and an ideal zero-temperature metal.  Here we ellucidate the issue, and give the formal definition of superfluid density in terms of response functions and the helicity modulus.  We will use this latter definition 
throughout the paper.

In the absence of impurities, metals are characterized by a resistivity which falls with temperature.  At zero temperature they support dissipationless currents.  The distinction with superfluidity is the robustness against adding disorder:  Weak disorder does not cause dissipation in a superfluid, but it does in a metal.

In dimensions $d>1$ superfluids and metals can be distinguished by the properties of the transverse current-current correlation function
% Characterizing, and distinguishing these systems requires considering the 
% The linear response of a superfluid is characterized by the superfluid density, while the equivalent quantity for the zero-temperature metal is the Drude weight.  
% Superfluids and metals are defined by the presence of zero-viscosity superflow and dissipationless currents, respectively (the latter is broadened by scattering at finite temperatures). While conceptually distinct, these features bear a certain phenomenological similarity that becomes relevant in one dimension. Quantitatively, it is useful to define a superfluid as a system that has a finite superfluid density (i.e. finite support for superflow) and a metal as a system that has a finite Drude weight (finite support for dissipationless current). In higher-dimensional systems, one can define both the superfluid density and the Drude weight in terms of a transverse current-current correlation function~\cite{sfCriterion}. The finite-temperature current-current correlation function is defined as
\begin{equation}
    T_{\alpha\alpha}({\bf q},i\omega_n)=\frac{1}{N}\int_0^\beta d\tau e^{i\omega_n\tau}\langle j_\alpha({\bf q},\tau)j_\alpha(-{\bf q},0)\rangle.
\end{equation}
Here 
$\omega_n=2\pi nT$ 
are the Matsubara frequencies 
and $\beta=1/T$ is the inverse temperature (we henceforth set $k_B=1$). This is the transverse correlation function when $\bf q$ is orthogonal to $\hat \alpha$.  
%${\bf q}\cdot\hat{\alpha}=0$ (which we now assume). 
The correlation function at real frequencies is obtained by analytic continuation $i\omega_n\to\omega+i\delta$.  {\em Note that this correlation function cannot be defined in one dimension as there is no transverse direction.}

In linear response theory, the current-current correlation function quantifies the amount of current generated by  a vector potential (or a fictitious vector potential which appears from moving frames). If we consider fluid flow in a pipe, the longitudinal response is typified by having moving end-caps, while the transverse response corresponds to moving an open pipe.  In a superfluid, only the normal component will move with the walls, and the superfluid density is given by
$\rho_s=\pi m^* D_s$, where $m^*$ is the effective mass and~\cite{sfCriterion}
\begin{equation}
    D_s=-\langle K_x\rangle- \lim_{q_y\to 0}T_{xx}(q_y,0)
    \label{eq:sfCC}.
\end{equation}
In the context of a Hubbard model, $\langle K_x\rangle$ is the expectation value of the kinetic energy per site due to motion in the $\hat{x}$ direction. This static response corresponds to the Hess-Fairbank effect previously introduced.  Note that $T_{xx}({\bf q},0)$ is poorly behaved at $\bf q=0$, as the longitudinal $T_{xx}(q_x,0)$ and transverse $T_{xx}(q_y,0)$ responses differ.

%In characterizing an ideal metal, the distinction between longitudinal and transverse response does not enter.  Instead, it 
By taking limits in a different way, one can calculate the Drude weight~\cite{sfCriterion}:
\begin{equation}
    D=-\langle K_x\rangle-\lim_{\omega\to0}T_{xx}({\bf 0},\omega)
    \label{eq:drudeCC}
\end{equation} 
This corresponds to the response to a homogeneous electric field.  Again, the limit is necessary as the point $(q=0,\omega=0)$ is singular.
In a superfluid, both $D_s$ and $D$ are non-zero, in a metal $D_s=0$ but $D\neq 0$, and in an insulator both $D_s$ and $D$ vanish~\cite{sfCriterion,mukerjeeShastry}.
% Note that we've assumed that the transverse momentum points only in the $\hat{y}$ direction. By linear response theory, this corresponds to the amount of particle current in the $\hat{x}$ direction which is generated by a vector potential $A_x({\bf q},\omega)$. The Drude weight can be defined in an analogous manner, simply reversing the order of limits~\cite{sfCriterion}:
While Eq.~(\ref{eq:sfCC}) is not well defined in one-dimension, Eq.~(\ref{eq:drudeCC}) is.  

%As already emphasized, momentum is a scalar in one dimension, so we cannot define a transverse current-current correlation function. This does not change our definition of the Drude weight, as Eq.~(\ref{eq:drudeCC}) does not rely on the correlation function being transverse, but it poses a problem for the superfluid density.

In order to extend the definition of superfluid density to one dimension, it is useful to reformulate the problem in terms of the helicity modulus~\cite{fisherHelicityModulus}. The helicity modulus, $\Upsilon$, gives the free energy response of the system to a twist of the boundary conditions. For example, if the $d$-dimensional many-body wavefunction
%in a space of length $L_\alpha$ along the direction $\hat{\alpha}$ 
obeys $\Psi({\bf x})=e^{i\Phi}\Psi({\bf x}+L_\alpha\hat{\alpha})$, then at finite temperature one defines
%can define the helicity modulus as
\begin{equation}
    \frac{1}{V}\big(F[\Phi]-F_0)=\frac{1}{2}\Upsilon\bigg(\frac{\Phi}{L_\alpha}\bigg)^2+\ldots
    \label{eq:helMod}
\end{equation}
where $V$ is the volume of the system, and  $F$ is the free energy, and $L_\alpha$ is the length of the system along $\hat{\alpha}$.  By the fluctuation dissipation theorem~\cite{fisherHelicityModulus}, 
$\Upsilon=\hbar^2\pi D_s$, and Eq.~(\ref{eq:helMod}) can be used to define the superfluid density.  Unfortunately, the $T\to0$ limit and the $L_\alpha\to\infty$ limits do not commute.  If one takes the $T\to0$ limit of Eq.~(\ref{eq:helMod}),
\begin{align} 
   \frac{1}{V}\big(E[\Phi]-E_0)=\frac{1}{2}\Upsilon_0\bigg(\frac{\Phi}{L_\alpha}\bigg)^2+\ldots\label{eq:gsHelMod}
\end{align}
then the helicity modulus instead gives the Drude weight, 
%, $E$ is the ground state energy, and $L_\alpha$ is the length of the system along $\hat{\alpha}$. Note that the application of a boundary condition twist is equivalent to threading the system with some magnetic flux. The helicity modulus and its ground state analog, $\Upsilon_0$, can be related to the superfluid density and Drude weight, respectively: $\Upsilon=\hbar^2\pi D_s$ and
$\Upsilon_0=\hbar^2\pi D$~\cite{fisherHelicityModulus}.
%Conceptual progress can be made by reformulating our definition of the superfluid density in terms of the change in free energy due to a twist in the boundary conditions~\cite{fisherHelicityModulus}. In the one dimensional case, this is identical to threading a ring (or any periodic direction in higher dimensions) with some magnetic flux. In terms of the boundary twist $\Phi$ along the $\hat{\alpha}$ direction, which means that the many-body wavefunction obeys $\Psi({\bf x})=e^{i\Phi}\Psi({\bf x}+L\hat{\alpha})$, the Drude weight and superfluid density are given by
% \begin{align}
%     D&=\frac{\pi L}{\hbar}\frac{d^2E[\Phi]}{d\Phi^2}\label{eq:drudeF}\\
%     D_s&=\frac{\pi L}{\hbar}\frac{d^2F[\Phi]}{d\Phi^2}
%     \label{eq:sfF}
% \end{align}
%where $E[\Phi]$ is the ground state energy and $F[\Phi]$ is the free energy.

There is no conceptual difficulty in extending the definitions in Eq.~(\ref{eq:helMod}) and (\ref{eq:gsHelMod}) to one dimensional systems, so this method succeeds in providing a consistent definition of superfluid density. In terms of the original formulation, 
%We have not succeeded in evading our quandry, however --
Eq.~(\ref{eq:helMod}) amounts to defining the superfluid density in terms of the $\omega=0$, $q\to 0$ limit of the (scalar) current-current correlation function~\cite{giamarchiShastry}.  If one reverses those limits (setting $q=0$ and taking the limit $\omega\to 0$) then one produces the Drude weight.  As will be argued below, there is a similar story involving the limits $T\to 0$ and $L\to\infty$. 
%It is by no means obvious that this is correct. 
%Clearly we find that the Drude weight and superfluid density can be distinguished at finite temperature.

\subsection{\label{sec:limits}Thermodynamic and zero temperature limits}
%On the basis of Eqs.~(\ref{eq:helMod}) and (\ref{eq:gsHelMod}), one might expect that $\Upsilon\to\Upsilon_0$ as the temperature $T\to 0$. This is not necessarily the case, however, as the thermodynamic limit does not in general commute with the zero-temperature limit~\cite{affleckQMC,svistunovSFDensity}. We can see this by determining the helicity modulus of a Luttinger liquid explicitly.
%As the Luttinger liquid Hamiltonian can be exactly diagonalized (see Eq.~(\ref{eq:diagHLL})), we can evaluate the helicity modulus explicitly. 
%It is quite clear from Eq.~(\ref{eq:hll}) that $\Upsilon_0=\hbar v_j/\pi$. As for the free energy, one finds~\cite{affleckQMC} that the ratio $\Upsilon(L,T)/\Upsilon_0$ is a scaling function of the product $LT$,
Comparing Eq.~(\ref{eq:hll}) with Eq.~(\ref{eq:gsHelMod}), the zero temperature helicity modulus is $\Upsilon_0=\hbar v_j/\pi$.  As first shown by Affleck \cite{affleckQMC}, one can calculate $\Upsilon$ by summing over states with all possible twists (see Appendix~\ref{sec:LLHelicityModulus}), finding
\begin{equation}
    \Upsilon(L,T)/\Upsilon_0=1+\frac{\pi^2\Upsilon_0}{LT}\frac{\vartheta_3^{\prime\prime}(0,e^{-2\pi^2\Upsilon_0/LT})}{\vartheta_3(0,e^{-2\pi^2\Upsilon_0/LT})},
    \label{eq:scalingFunc}
\end{equation}
where $\theta_3(z,q)$ is the Jacobi theta function of the third kind. The $T\to0$ and $L\to\infty$ limits do not commute: $\Upsilon/\Upsilon_0$ approaches $1$ as $LT\to 0$ and approaches $0$ as $LT\to\infty$. Taking the thermodynamic limit prior to $T\to 0$ results in a formally vanishing superfluid density. 

This structure arises from a competition between the thermal energy-scale $T$ and the gaps  between states in different topological sectors: 
%the broad Gibbs distribution of states with finite current: 
a many-body state with a $2\pi$ phase twist across its length $L$ (which consequently supports finite current) differs in energy from the ground state by $\Delta E_{2\pi}=2\pi\hbar v_j/L$.
When one takes the temperature to zero in a system with finite $L$, one only occupies states with a fixed winding, resulting in a superfluid response. The opposite limit yields a large ensemble of windings, and the system behaves like a normal fluid.
%Clearly this energy difference vanishes as $L\to\infty$. Thus, a Gibbs distribution which includes all states within $k_BT$ of the ground state will sample all such current states. This causes the expectation value of the equilibrium current operator to vanish. 
Note that this peculiarity is specific to 1D: one finds $\Delta E_{2\pi}\propto L^{d-2}$ in $d$ spatial dimensions, so the energy gap will be finite in the thermodynamic limit for $d\geq 2$~\cite{svistunovSFDensity}.
%For $d$ spatial dimensions, one finds $\Delta E_{2\pi}\propto L^{d-2}$ -- thus, it is only when $d=1$ that this energy cost vanishes in the thermodynamic limit~\cite{svistunovSFDensity}.

In our calculation we explicitly work at $T=0$, and these considerations are irrelevant:  our procedure correctly yields $\Upsilon_0$ and hence the superfluid density.

\subsection{\label{sec:neq}Non-equilibrium considerations}
The arguments so far have been thermodynamic in nature and assumed thermal equilibrium. The energy barriers separating topologically distinct sectors do not vanish in the thermodynamic limit.
Therefore the time to equilibrate will be exponential in $1/T$, even though the states with different windings have degenerate energies.
%, leading to exponentially long relaxation times.  
These long relaxation times must be taken into account in modeling 
experiments in
cold atom systems~\cite{coldAtomSFExpt1,coldAtomSFExpt2} and $^4{\rm He}$ nanopores~\cite{heSFExpt1,heSFExpt2,heSFExpt3,heSFExpt4}. One approach is to introduce   
% have called into doubt the physical relevance of the equilibrium calculation. In particular, these experiments find that the scaling function in Eq.~(\ref{eq:scalingFunc}) does not describe the length-dependence of the measured superfluid density. This has lead to the development of 
``dynamical"  superfluidity~\cite{cazalillaDynamicalSF,danshitaPhaseSlips,hcBosonsPhaseSlips,xyModelPhaseSlips}. 

%The equilibrium ensemble contains many nearly degenerate states, each carrying different amounts of current.  These are, however, separated by barriers which do not vanish in the thermodynamic limit.  Hence the equilibration timescale will diverge at low temperatures.  The experimentally relevant ensemble  therefore involves a time-scale dependent subset of these states.

There is a close connection between this dynamical superfluidity and the physics described in Sec.~\ref{sec:limits}.  In the equilibrium theory, taking $T=0$ then the limit $L\to \infty$ freezes the system into a single current-carrying sector and yields a finite superfluid density, analogous to the dynamical superfluid density.  The opposite limit yields no phase stiffness.  The theory of dynamical superfluid density generalizes this argument to predict the temperature dependent response of the experimental system.

While this nonequilibrium physics can be very important, we will simply focus on equilibrium superfluidity at zero temperature.

\section{\label{sec:methods}Methods}
We compute the ground state phase diagram of the 1D Bose-Hubbard model using two infinite tensor network algorithms: iDMRG~\cite{whiteDMRG1,whiteDMRG2,mcculloch2008infinite} and VUMPS~\cite{vumps,vumps2}. We make use of the ITensor library~\cite{itensor} in our implementations. In this section we discuss the relevant features of these techniques as well as our approach to computing the superfluid density. We provide a detailed discussion of the VUMPS algorithm in Appendix~\ref{sec:vumps}.

%\subsection{iDMRG}
Both iDMRG and VUMPS are variational techniques that make use of a matrix product state ansatz:  As a basis for the many-body state one considers states with a fixed number of bosons on each site $\{n_j\}$;  in the thermodynamic limit $j$ runs from $-\infty$ to $\infty$. The wavefunction in this basis 
%$\psi(\cdots,n_1,n_2,\cdots)$.  One writes this coefficient as 
is written as a product of matrices, 
\begin{equation}
\psi(\cdots,n_1,n_2,\cdots)= 
\sum_{\{s\}}
%{\cdots s_1,s_2\cdots}
\cdots { A}^{n_1}_{s_0s_1} { B}^{n_2}_{s_1s_2}{ C}^{n_3}_{s_2s_3}\cdots
\end{equation}
where the sum over $\{s\}$ represents all possible values of the ``bond indices" $s_j$.
The number of values that
each $s_j$ takes on is referred to as the bond dimension $\chi$.  Describing states with more entanglement requires larger $\chi$. An arbitrary state can be written in this form if the bond dimension is sufficiently large. Both iDMRG and VUMPS find the lowest energy matrix product state with some enforced constraints on the bond dimension. They principally differ in how they carry out the minimization.

%find the lowest energy 
%state 
%The infinite density matrix renormalization group (iDMRG) algorithm minimizes the expectation value of a Hamiltonian over the set of infinite 
%infinite matrix product state~\cite{whiteDMRG1,whiteDMRG2}.

%While tensor network methods have been employed for a variety of uses (cite), their original~\cite{whiteDMRG1,whiteDMRG2} and most common purpose is to minimize the expectation value of a given Hamiltonian over the set of matrix product states. 
%Such matrix product states comprise a class of variational ansatze which efficiently model quantum states in which the entanglement between two regions is small.  The degree of entanglement that can be captured is controlled by the "bond dimension" -- larger bond dimension states can encode more entanglement, at the expense of more computational effort.

%y are characterized by a parameter,
% the bond dimension, that quantifies the maximum entanglement allowed between neighboring sites. Matrix product states with large bond dimensions have been used to provide accurate numerical results for a variety of models, and the range of techniques used to optimize them has grown considerably in recent years (see Ref.~\cite{SCHOLLWOCKreview} for a review). 

\begin{figure}
    \centering
    \includegraphics[width=3.375in]{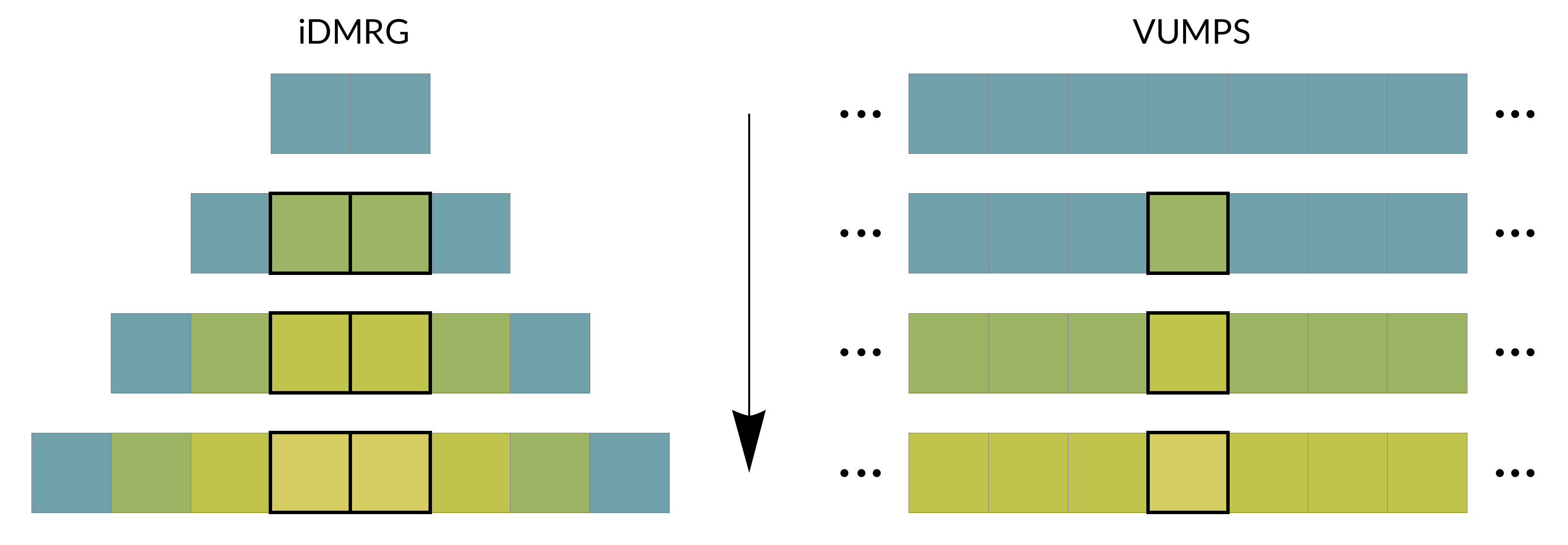}
    \caption{Schematic showing the difference between the iDMRG and VUMPS algorithms. Blocks denote a matrix product state composed of the contraction of single-site tensors. In each iteration, the state from the previous iteration serves as a ``bath" from which the next optimal state is chosen. While iDMRG grows a finite chain outwards, VUMPS performs global updates after each iteration and enforces that the state be translationally-invariant.}
    \label{fig:schematic}
\end{figure}

%The iDMRG algorithm is a method for computing the optimal translationally-invariant matrix product state, thereby approximating the ground state in the thermodynamic limit~\cite{whiteDMRG2}. 
The iDMRG algorithm begins by choosing an initial two-site MPS. For example, one could start with the exact ground state of the two-site problem written as an MPS: 
$\psi_0(n_1,n_2)=\sum_s A^{n_1}_s Z^{n_2}_s$.  In Fig.~\ref{fig:schematic} this initial state is depicted as two boxes, representing $\bf A$ and $\bf Z$. After truncating the bond dimension and appropriately normalizing the matrices, one appends two sites to the center of the chain, finding matrices $\bf B$ and $\bf Y$ which minimize the energy  of the four-site problem with
$\psi_1(n_1,n_2,n_3,n_4)=\sum_{\{s\}} \bar A^{n_1}_{s_1} B^{n_2}_{s_1 s_2} Y^{n_3}_{s_2 s_3} \bar Z^{n_4}_{s_3}$.  Here $\bf \bar A$ and $\bf \bar Z$ are the transformed versions of $\bf A$ and $\bf Z$, and are held fixed during the optimization with respect to $\bf B$ and $\bf Y$ \cite{SCHOLLWOCKreview}. 
%. The two-site MPS added to the center is chosen such that the energy of the four-site chain is minimized, given the constraint that the outer two sites are fixed.
As depicted in the figure, this procedure is iterated until the matrices added to the center in successive iterations are sufficiently similar. One then approximates the translationally-invariant ground state of the Hamiltonian as an infinite chain composed of those matrices.  
%This two-site state updating procedure 
%(that is, the appending of two sites to the center of the chain instead of one) 
%allows one to update the bond dimension after each iteration. By performing an SVD on the central two sites, one can truncate the least important bonds up to some overall tolerance. In practice one picks a maximum allowed bond dimension as well as a tolerance for each run. Increasing the bond dimension increases the accuracy of the variational ground state properties, but it also increases the size of the eigenvalue problem that must be solved to append the central sites. Notably, even when one performs successive updates in which the bond dimension does not change, the central two-site tensor is still truncated in each iteration. This means that iDMRG has residual truncation error that does not vanish even at its fixed point. This truncation error will vanish in the infinite bond dimension limit, so extrapolation is necessary to achieve numerically accurate results.

%\subsection{VUMPS}
%The VUMPS algorithm has been recently developed~\cite{vumps,vumps2} as a more efficient method for finding the optimal translationally-invariant matrix product state approximation of the ground state. 
%Rather than iteratively growing a finite-length matrix product state, 
The iterative growth procedure of iDMRG can be compared to VUMPS, where, as illustrated in Fig.~\ref{fig:schematic}, a single site is inserted in the middle of an infinite matrix product state.  One finds the tensor for that site that minimizes the energy and then constructs an infinite product state from it.
%the VUMPS algorithm globally updates an infinite-length MPS after each iteration. One then performs a single-site update, inserting a central tensor and minimizing the energy with respect to the surrounding infinite system, as shown schematically in Fig.~\ref{fig:schematic}. 
As we will discuss in more detail in Sec.~\ref{sec:results},  this global update is particularly useful when the ground state has long-range correlations and allows one to overcome some of the bottlenecks present in the iDMRG algorithm's local updates.
%as is the case in the Luttinger liquid phase of the 1D Bose-Hubbard model, 
%because the VUMPS algorithm computes the energy in the thermodynamic limit after each iteration.
%This method should be particularly useful when the ground state entanglement entropy diverges with system size, as is the case in the Luttinger liquid phase of the 1D Bose-Hubbard model. The original VUMPS algorithm, proposed in Ref.~\cite{vumps}, utilizes a single-site updating procedure where the bond dimension of the matrix product state remains fixed. 

There are some additional technical differences between our implementations of iDMRG and VUMPS which are related to single-site vs. two-site updating.  
%[There exist single-site iDMRG algorithms and 2-site VUMPS algorithms.]
In a single-site update procedure  one finds a true variational minimum at fixed bond dimension, while in a two-site approach there is a truncation error associated with decomposing the two sites~\cite{whiteOneSiteDMRG,vumps}. Conversely, the two-site procedure samples a larger variational subspace and more readily allows for dynamically changing the bond dimension.  Although we do not report the results here, we also implemented a 2-site VUMPS algorithm.  We found that the truncation error interacted poorly with the procedure of constructing the infinite matrix product state, resulting in less accurate results for the same bond dimension. 

%The single-site updating procedure means that the bond dimension remains fixed from iteration to iteration.
%This can be a practical shortcoming, so when necessary we develop a way to dynamically change the bond dimension. This is detailed in Appendix~\ref{sec:vumpsDetails}. 
%The benefit of single-site updates, however, is that the updates are performed without any truncation error. This means that the converged state at fixed bond dimension will be the true variational minimum, unlike iDMRG.

%We develop a variant of the original algorithm that utilizes a two-site updating procedure, akin to what is used in the original DMRG and iDMRG algorithms~\cite{whiteDMRG1,whiteDMRG2,SCHOLLWOCKreview}. This is detailed in Appendix~\ref{sec:vumpsDetails}. The benefit of this updating procedure is that the bond dimension is varied after each iteration, which can improve efficiency. A natural consequence of our construction is that we optimize over translationally-invariant matrix product states with a two-site unit cell. This allows us to capture symmetry-broken phases, such as charge density waves, at lower bond dimensions. It is, however, worth noting that two-site unit cells can be accommodated by the original VUMPS algorithm~\cite{vumps,vumps2}.

\subsection{\label{sec:calcSF}Calculating $\rho_{s}$}
%We calculate the superfluid density, $\rho_s$, by computing the helicity modulus $\Upsilon_0$ that is defined in Section~\ref{sec:definitions}. Rather than applying a boundary condition, 
We calculate the superfluid density by first applying a gauge transformation
%transforming the Hamiltonian so that a phase $\phi$ accumulates across neighboring sites, 
$Ua_jU^\dagger=e^{-i\varphi j}a_j$ to the terms in the Hamiltonian. We then construct the lowest-energy \textit{uniform} matrix product state.
This results in a current carrying state and is  analogous to having twisted boundary conditions~\cite{shastryTBC}. The helicity modulus is extracted from the energy as a function of $\varphi$ according to Eq.~(\ref{eq:gsHelMod}).  The superfluid density is then $\rho_s=\Upsilon_0/(2 t)$, where $t$ is the hopping matrix element in Eq.~(\ref{eq:Hbh}).

We emphasize that this procedure is not the same as simply applying a gauge transformation to the ground-state wavefunction.  The gauge transformation is not homogeneous, and hence converts a uniform matrix product state to a non-uniform one.  The tensors in our wavefunction can be used to make a length $L$ matrix product state on a ring with a phase twist $\Phi=L \varphi$ across the boundary.

\section{\label{sec:results}Results}
As reviewed in Sec.~\ref{sec:LL},
the Luttinger liquid phase is critical,  with an infinite correlation length and  power-law decaying correlation functions [see Eq.~(\ref{eq:dMatLL})]. Consequently the entanglement entropy diverges.
%An MPS ansatz can only capture this infinite correlation length in the limit that the bond dimension $\chi\to\infty$. 
%This structure is evident in the behavior of various quantities with bond dimension.
An MPS with finite bond dimension will be an approximant, with finite entanglement entropy.
%, and artificially cut-off correlation functions.
The critical stucture can be revealed by studying how various quantities scale with bond dimension.  Such 
%At finite bond dimensions, one can determine this critical structure by studying how various quantities rescale with bond dimension. This is interpreted most naturally as 
finite entanglement scaling~\cite{CalabreseLefevre}
%as the fixed-$\chi$ approximation literally truncates the entanglement spectrum, but it can also be understood as a 
is closely related to finite-size scaling,
%finite-length scaling when the MPS correlation length is sufficiently large
where the bond-dimension is viewed as a control parameter which adjusts a spatial cut-off~\cite{jMooreEE}. 
%In what follows, both of these interpretations will emerge. 

Local quantities (energy, short range correlations, etc.) converge rapidly with bond dimension.  Long-range properties are readily found using scaling analysis.  As described below, one sees excellent scaling collapse with moderate bond dimensions: $\chi\sim 20-50$.  
%As explained in Sec.~??, the superfluid density can be extracted from the twist-dependence of the energy, and hence only requires moderate $\chi$.

%While the MPS ansatz at finite bond dimension is not well-suited to capturing the behavior of a critical phase, we note that once the critical properties can be identified by a scaling collapse, there is nothing more to be learned by moving to higher bond dimension. As we know the structure of the low-energy theory, the scaling collapse indicates that we are in a regime where we can extract the Luttinger parameters. Optimizing the MPS ansatz at higher bond dimensions will only improve the long-wavelength behavior of the ansatz, which can already be inferred from the scaling function.

%{\color{red}  Do we want to foreshadow any of the calculations involving interplay  between  local and long-range physics?}

In Sec.~\ref{sec:densityMatrix} we show the behavior of the single-particle density matrix and define the correlation length. We also compare the convergence properties of iDMRG and VUMPS in the superfluid phase, attributing the superiority of the latter to finite-size effects in the iDMRG algorithm. In Sec.~\ref{sec:momentum} we discuss the properties of the momentum distribution and demonstrate finite entanglement scaling via a scaling collapse. In Sec.~\ref{sec:sfResults} we plot the superfluid density across the phase diagram and discuss its relationship to the single-particle density matrix. We also determine the Luttinger parameter, $K$, as a function of $\mu/U$ and $t/U$. In Sec.~\ref{sec:entanglement}, we conclude by discussing how the entanglement of the MPS ansatz scales with bond dimension, extracting the conformal exponent $\kappa$ predicted in Ref.~\cite{jMooreEE}.

\subsection{\label{sec:densityMatrix}Density matrix: iDMRG and VUMPS}

%that depends on the Luttinger parameter (see Eq.~(\ref{eq:dMatLL})). 
%As Luttinger liquid theory describes the 1D Bose-Hubbard model at low energies and long wavelengths, this result will only hold for large spatial separations.

\begin{figure}
    \centering
    \includegraphics[width=3.375in]{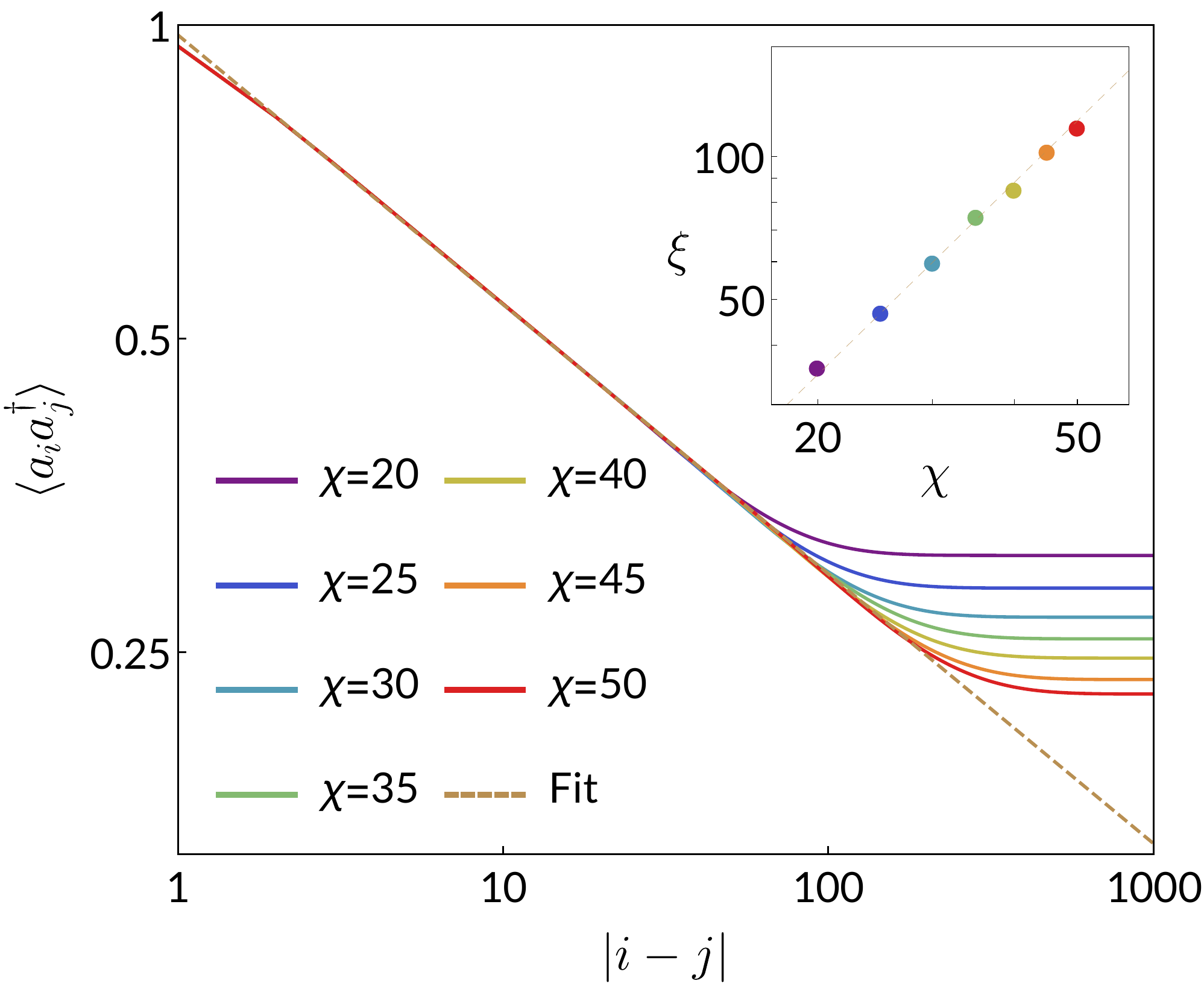}
    \caption{Log-log plot of the density matrix, $\langle a_ia^\dagger_j\rangle$, versus spatial separation, $|i-j|$, for a variety of bond dimensions. Data is taken at the point $(t/U,\mu/U)=(0.2,0.5)$. 
    The Luttinger parameter is extracted from the slope of the 
    power-law region (dashed line). %is found over a length-scale which grows with $\chi$.
    %Intermediate-range data points display a sizeable region of algebraic decay. We extract the Luttinger parameter by fitting the slope of the decay, shown by the dashed line. 
    (inset) Plot of the correlation length, $\xi(\chi)$, computed using Eq.~(\ref{eq:xi}), versus bond dimension on a log-log scale. Also shown is a fit (dashed line) of the form $\xi(\chi)=\alpha \chi^{\kappa}$ with $\kappa=6/(1+\sqrt{12})$ and $\alpha$ a free parameter.}
    \label{fig:densityMatFig}
\end{figure}

Figure~\ref{fig:densityMatFig} shows the single particle density matrix $\langle a_ia^\dagger_j\rangle$ as a function of spatial separation $|i-j|$ for 
%Luttinger liquid  theory (Sec.~\ref{sec:LL}) predicts that for large separations $|i-j|$, the density matrix, $\langle a_ia^\dagger_j\rangle$, should fall off  as $|i-j|^{-K/2}$.
%A log-log plot of the density matrix, computed for various bond dimensions, $\chi$, is shown in Fig.~\ref{fig:densityMatFig}. The data is taken at
a representative point in the superfluid phase, $(t/U,\mu/U)=(0.2,0.5)$. 
The expected Luttinger liquid algebraic decay is seen  over a wide range of separations.  
The finite bond dimension introduces an artificial cutoff beyond which $\langle a_ia^\dagger_j\rangle$ is constant.
It is natural to define a bond-dimension-dependent quasicondensate density
\begin{equation}
    \lim_{|i-j|\to\infty}\langle a_ia^\dagger_j\rangle\to|\langle a_i\rangle|^2\equiv \rho_{qc}.
\end{equation}
The fact that there is no Bose-Einstein condensation in 1D is manifest in the fact that $\rho_{qc}\to0$ as $\chi\to\infty$.  The correlation length, defined by
\begin{equation}
    \xi^2(\chi)=\bigg(\sum_j j^2\langle a_0a_j^\dagger\rangle_c\bigg)/\bigg(\sum_j\langle a_0a_j^\dagger\rangle_c\bigg),
    \label{eq:xi}
\end{equation}
is shown in the inset of Fig.~\ref{fig:densityMatFig}. 
Here $\langle a_0a_j^\dagger\rangle_c=
\langle a_0a_j^\dagger\rangle-\rho_{qc}$.
The correlation length grows as a power law, $\xi(\chi)\propto \chi^\kappa$, where $\kappa=6/(1+\sqrt{12})$, as expected from the conformal invariance of the Luttinger liquid~\cite{CalabreseLefevre,jMooreEE}.
 
This correlation length can be viewed as a spatial cutoff, and the quasicondensate in this model is very similar to quasicondensates found in finite length systems~\cite{1D_boson_review}. In Appendix~\ref{sec:quasicondensate} we show that  
$\rho_{qc}\sim\xi^{- K/2}$,
which is similar to the finite-size scaling
$\rho_{qc}\sim L^{- K/2}$ in Refs.~\cite{shlyapnikovQuasiCond,GangardtQuasiCond,rigolQuasiCond}.

\begin{figure}
    \centering
    \includegraphics[width=3.375in]{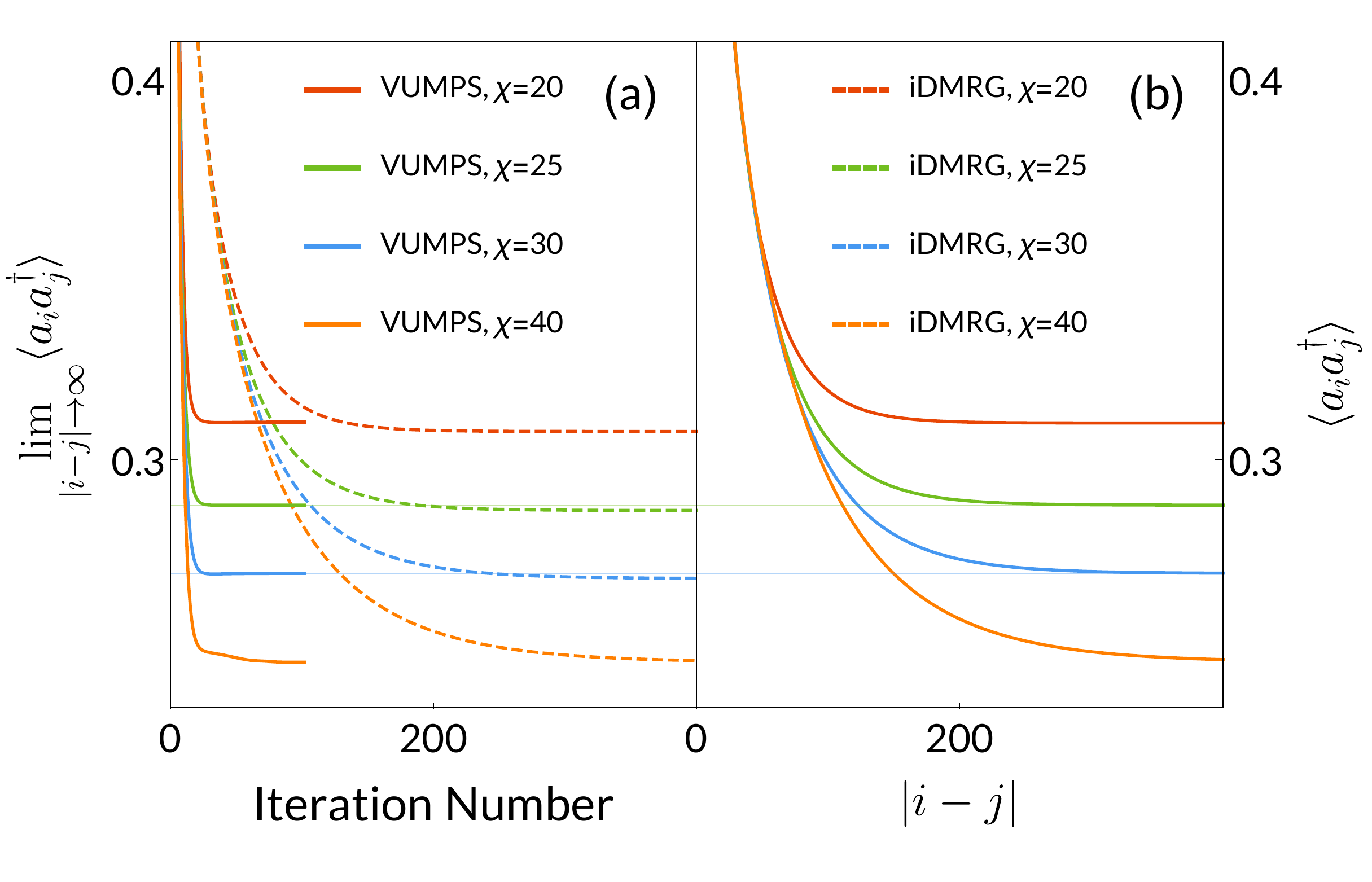}
    \caption{{(a)} Plot of the quasicondensate density, $\lim_{|i-j|\to\infty}\langle a_ia^\dagger_j\rangle$, versus iteration number for a single run of VUMPS (solid) and iDMRG (dashed). Results are plotted for various bond dimensions at the point $(t/U,\mu/U)=(0.2,0.5)$. Convergence can be understood when the curves saturate at the asymptotic limits (translucent lines). Clearly VUMPS converges in fewer iterations than iDMRG. The discrepancy between the asymptotic limits is due to the truncation error inherent in the two-site iDMRG algorithm; the one-site updates in VUMPS, by contrast, converge to the variational minimum. {(b)} Plot of the density matrix, $\langle a_ia^\dagger_j\rangle$, versus spatial separation $|i-j|$. Results are from fully converged states, computed with VUMPS, at the same point in parameter space. The decay of the density matrix with distance quantitatively mirrors the decrease in the quasicondensate density with iteration number in the iDMRG algorithm.}
    \label{fig:comparisonFig}
\end{figure}

\begin{figure*}[t]
    \centering
    \includegraphics[width=7in]{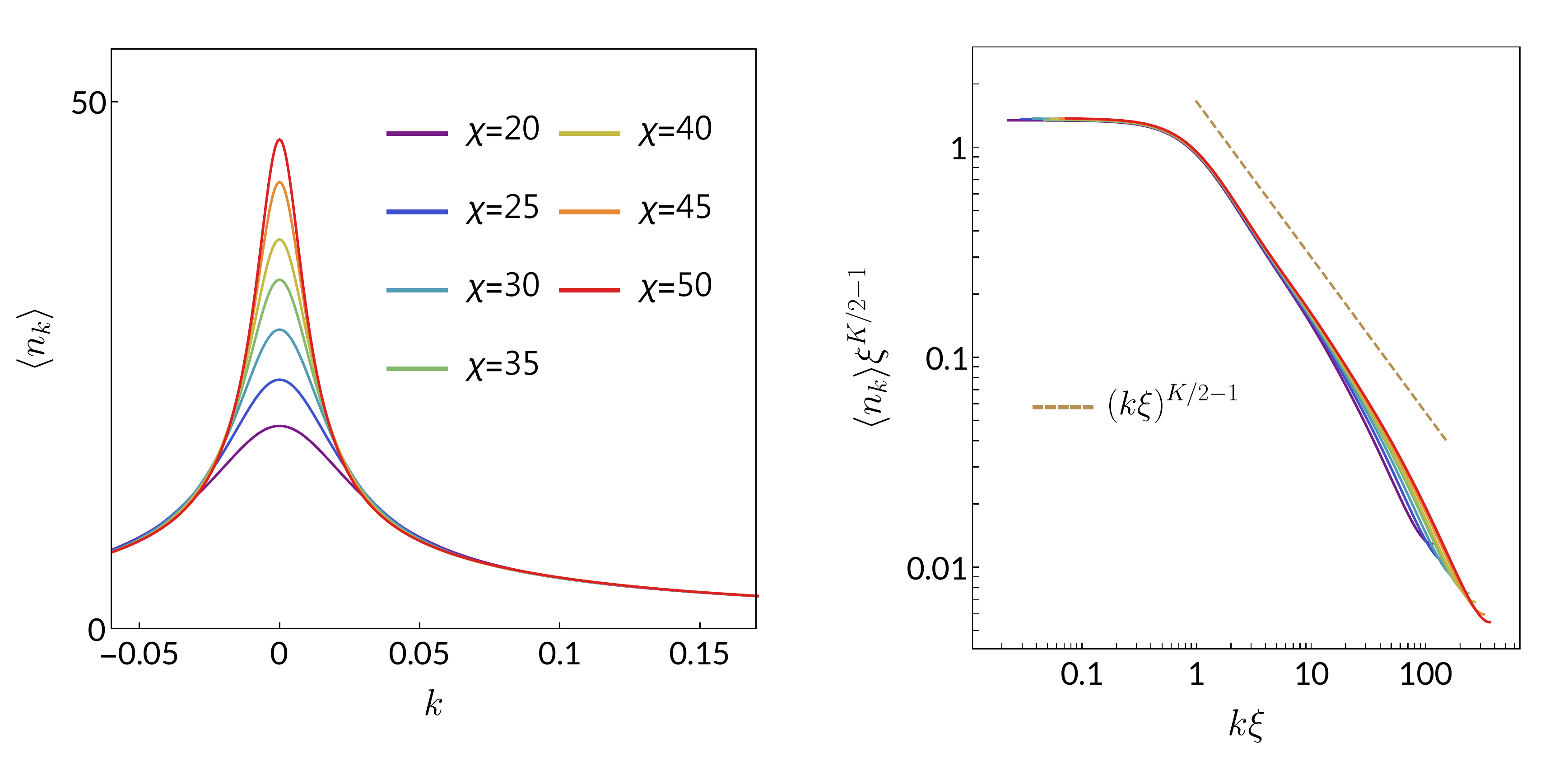}
    \caption{(a) Plot of the momentum distribution, $\langle n_k\rangle$, at the point $(t/U,\mu/U)=(0.2,0.5)$ for various bond dimensions. The divergence at $k=0$ is smoothly cut off at fixed bond dimension due to the finite correlation length. Note the curves are almost identical for $|k|>0.1$ in units of the reciprocal lattice constant. (b) Plot of the momentum distribution functions on a log-log scale after rescaling by powers of the correlation length. Note that the Luttinger parameter, $K$, was determined from a fit to the single-particle density matrix (see Fig.~\ref{fig:densityMatFig}). This captures the critical behavior for small momenta.}
    \label{fig:momentumFig}
\end{figure*}

%The interplay between finite-length and finite-entanglement simulations of a Luttinger liquid sheds 
These properties of the single particle density matrix shed
light on the convergence properties of iDMRG and VUMPS. In Fig.~\ref{fig:comparisonFig}(a) we plot the quasicondensate density, $\rho_{qc}$, versus iteration number for a single run of VUMPS (solid) and iDMRG (dashed) at various bond dimensions.   The quansicondensate density falls with the iteration number, eventually converging to a bond-dimension dependent constant.  
%Convergence of the state can be qualitatively judged by the curves saturating the asymptotic quasicondensate density, shown in translucent solid lines. 
%The first thing to note is that VUMPS and iDMRG 
Finite truncation error in the two-site state updates in the iDMRG algorithm limit its accuracy, leading to a slightly different value of $\rho_{qc}$ compared to VUMPS.
%do not saturate at the same values. This is due to finite truncation error in the two-site state updates in the iDMRG algorithm, which prevents iDMRG from converging to the variational minimum. By contrast, t
The one-site state updates used by VUMPS work at fixed bond dimension and hence do not introduce any truncation error. In addition to being more accurate, VUMPS converges in many fewer iterations than iDMRG. For $\chi=40$, a single iteration of VUMPS takes roughly twice as much computer time as a single iteration of iDMRG, and is therefore more efficient.
%We emphasize that, for the data points shown, iDMRG has already reached the maximum bond dimension -- thus, the discrepancy does not derive from two-site versus one-site updates. Rather, 

In Fig.~\ref{fig:comparisonFig}(b)
we show the spatial dependence of the converged density matrix,
%It is udeful to can be understood by comparing the iDMRG data on the left panel of Fig.~\ref{fig:comparisonFig} to the right panel, which shows the density matrix versus spatial separation. 
%The converged correlation function 
$\langle a_ia^\dagger_j\rangle$.   That correlation function, with $|i-j|=n$, is 
remarkably similar to
the long range correlations
$\rho_{qc}=\lim_{|i-j|\to \infty} \langle a_ia^\dagger_j\rangle$ of the $n$'th iteration of the iDMRG algorithm.
%the long-range correlations found in intermediate states of the iDMRG algorithm   Fig.~\ref{fig:comparisonFig} (a), with iteration number and distance playing equivalent roles. 
This structure is understood by noting that after $n$ iterations, iDMRG describes a system of length $2n$.  When $n<\xi$, this finite size 
introduces a cutoff. 
%of the matrix product state at each intermediate iteration of the iDMRG algorithm is what limits its convergence: just as $\langle a_ia^\dagger_j\rangle$ only saturates to $\rho_{qc}$ for $|i-j|\gtrsim \xi$, the intermediate MPS can only capture the asymptotic quasicondensate density when it is longer than $\xi$. The length of the MPS is equal to twice its iteration number, but seeing as states are only updated at the center of the chain, the relevant length scale is half its length. Thus, the number of iterations for iDMRG to converge scales linearly with the correlation length of the system. 
One consequence is that
%For large bond dimensions, that means that 
the number of iterations required for iDMRG convergence grows at least as fast as
$\xi\propto \chi^\kappa$.  VUMPS does not suffer this problem, and has better scaling with $\chi$.  This benefit should be found in any critical or gapless phase/point.
%only be achieved for some number of iterations $N(\chi)$ that scales as $N(\chi)\propto \chi^\alpha$ with $\alpha\geq\kappa$. While VUMPS will also require more iterations to converge at larger bond dimension, its scaling is not fixed by any geometric features of the algorithm. This illustrates the advantage of VUMPS over iDMRG in the Luttinger liquid phase, and in critical or gapless phases more broadly.

\subsection{\label{sec:momentum}Momentum distribution}
The non-condensed momentum distribution function, $\langle n_k\rangle$, is easily obtained as the Fourier transform of the density matrix:
\begin{equation}
    \langle n_k\rangle=\sum_j e^{ikj}\langle a_0a^\dagger_j\rangle_c,
    \label{eq:momentumDist}
\end{equation}
where as before, $\langle a_0a^\dagger_j\rangle_c=\langle a_0a^\dagger_j\rangle-\rho_{qc}$.
%The background quasicondensate density leads to a delta function peak at $k=0$. As this is a finite-correlation-length effect, in what follows we remove the peak by replacing $\langle a_0a^\dagger_j\rangle\to\langle a_0a^\dagger_j\rangle_c$ in Eq.~(\ref{eq:momentumDist}).
%We show that $\rho_{qc}$ vanishes as a power law with bond dimension in Appendix~\ref{sec:quasicondensate}, and thus can be ignored.

We plot $\langle n_k\rangle$ versus $k$ for a variety of bond dimensions at $(t/U,\mu/U)=(0.2,0.5)$ in panel (a) of Fig.~\ref{fig:momentumFig}. The momentum distribution function is sharply peaked about $k=0$.  This is not a signature of Bose-Einstein condensation, but is instead 
%. Rather, this feature is 
indicative of the critical scaling of the density matrix. At long distances the density matrix falls off as $r^{-K/2}$;
by power-law counting its Fourier transform scales as $k^{K/2-1}$ for small momenta.
%power law counting indicates that the Fourier transform of a quantity that vanishes as $r^{-K/2}$ at large distances in real space will diverge as $k^{K/2-1}$ for small momenta. 
As seen in the figure, this small-$k$ divergence is cut off %at fixed bond dimension due to 
by the finite correlation length in our matrix product state ansatz.  The correlation length grows with bond dimension, and the momentum distribution function approaches a power law as $\chi\to\infty$.
%which is found at finite bond dimension.
%of the uniform MPS, and thus the peak at $k=0$ grows with increasing bond dimension. 
For $k\gtrsim0.1$, $n_k$ is independent of $\chi$.  This is equivalent to the collapse in Fig.~\ref{fig:densityMatFig}, and indicates that the short-distance correlations are well-captured by a MPS with moderate bond dimension.
%s, as indicated by the lack of any scaling behavior for $k\gtrsim0.1$. 
In Fig.~\ref{fig:momentumFig}(b), we demonstrate a scaling collapse by rescaling the momentum and the distribution function by powers of the correlation length, $\xi(\chi)$, computed with Eq.~(\ref{eq:xi}). The asymptotic power-law behavior, indicated by the dashed line, is visible for $k\xi\sim10$. Around $k\xi\sim1$, that divergence is smoothly cut off and all curves approach a constant.
%where $\langle n_k\rangle\xi^{K/2-1}\sim\mathcal{O}(1)$.

\subsection{\label{sec:sfResults}Superfluid density}
\begin{figure}
    \centering
    \includegraphics[width=3.375in]{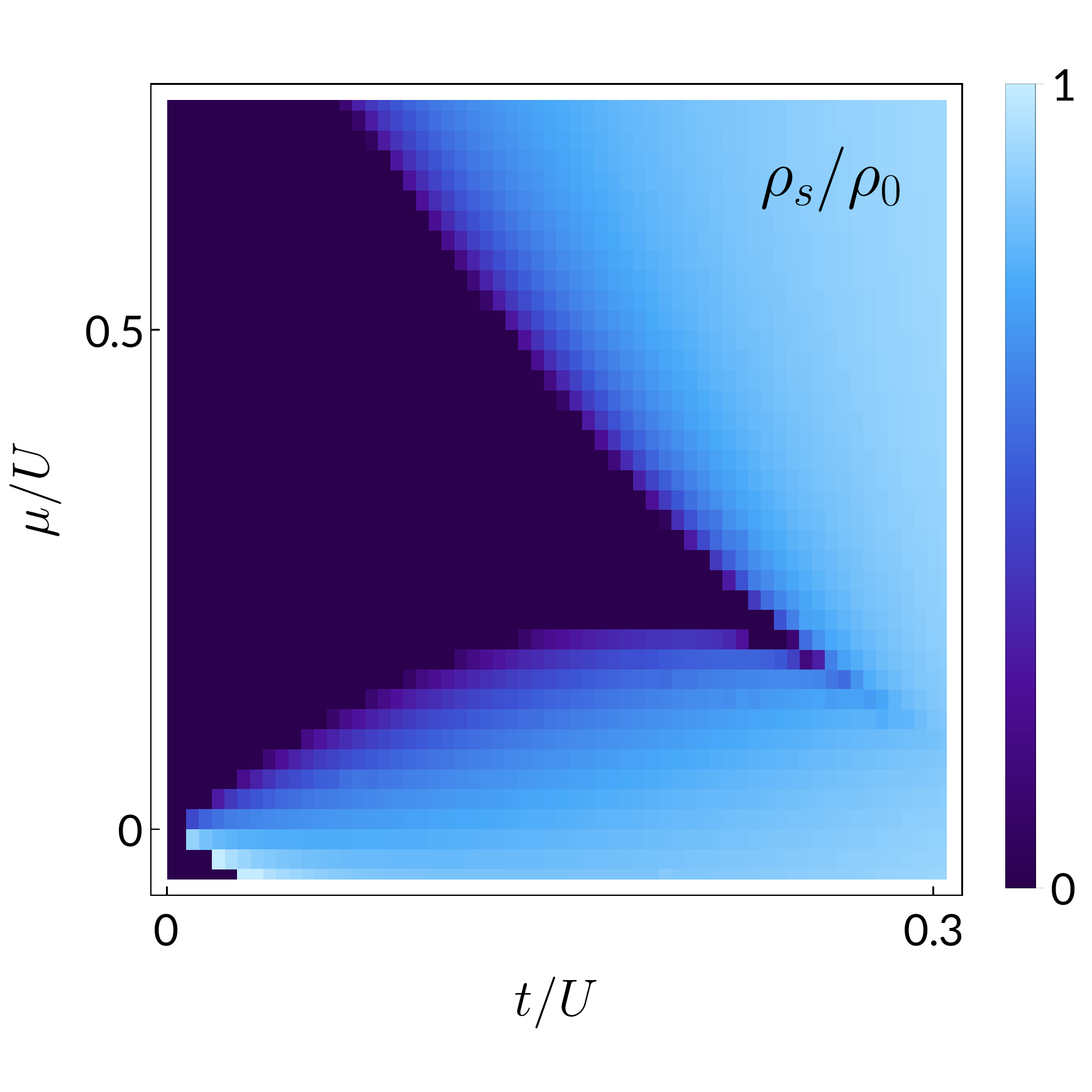}
    \caption{Superfluid fraction, $\rho_s/\rho_0$, as a function of $t/U$ and $\mu/U$ with $\chi=25$. For $\mu>0$, the dark region (indicating $\rho_s\to 0$) is the $n=1$ Mott lobe. For $\mu<0$, the dark region indicates the vacuum ($n=0$).}
    \label{fig:heatMap}
\end{figure}

As discussed in the introduction, unlike in a Galilean-invariant system, the zero-temperature superfluid fraction of the 1D Bose-Hubbard model continuously interpolates between 0 and 1. We plot the superfluid fraction, $\rho_s/\rho_0$, with bond dimension $\chi=25$ as a function of $t/U$ and $\mu/U$ in Fig.~\ref{fig:heatMap}. The $n=1$ Mott lobe is clearly visible as the dark region where the superfluid fraction vanishes. For $\mu<0$, the dark region indicates the vacuum. The VUMPS algorithm works directly in the thermodynamic limit and correctly captures the critical behavior away from the tip of the Mott lobe.  At the tip the 
%we see a mean-field-like transition at the Mott-superfluid boundary. The exception to this is at the Mott lobe tip, where 
transition is BKT-like, with an expected universal jump in the superfluid density~\cite{BKT1,BKT2,BKT3}.  This jump is rounded over at finite $\chi$.  Scaling analysis, however, can be used to locate the phase boundary.

%Such an 
%The BKT transition is an 
%infinite-order phase transition~\cite{BKT1,BKT2,BKT3}, so finite-correlation-length effects obscure the precise location of the tip.

%In three-dimensional strongly-correlated bosonic systems, the condensate fraction is related to the long-distance behavior of $\langle a_ia^\dagger_j\rangle$ while the superfluid fraction is generally unrelated to the density matrix. In one dimension, by contrast, the condensate fraction is formally zero while the superfluid fraction is related to the Luttinger parameter and hence the long-distance algebraic scaling of $\langle a_ia^\dagger_j\rangle$. 

\begin{figure}
    \centering
    \includegraphics[width=3.375in]{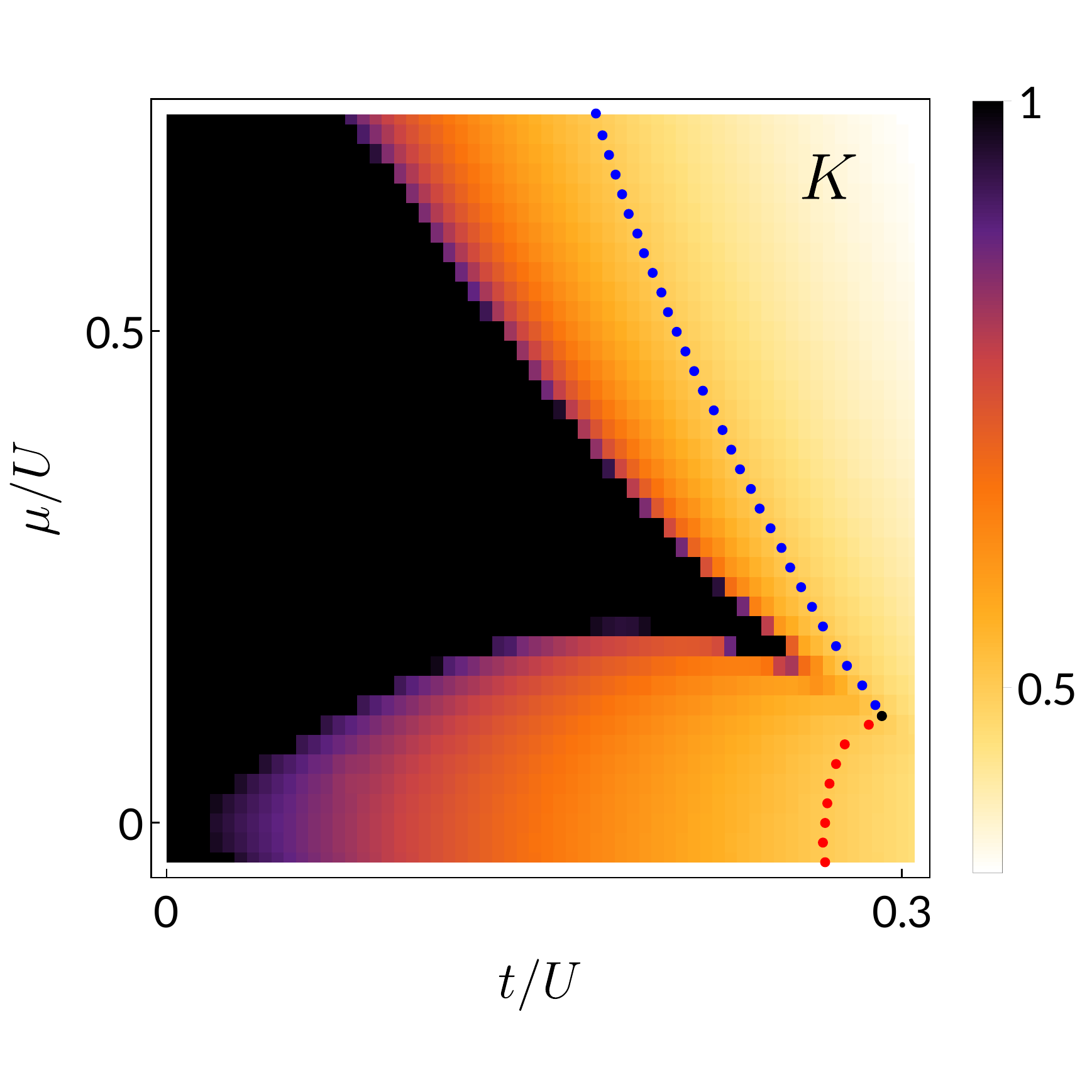}
    \caption{Luttinger parameter, $K$, as a function of $t/U$ and $\mu/U$ with $\chi=25$. In the superfluid phase, $0<K<1$; at the density-driven Mott-SF transition, $K\to 1$. The Mott lobe tip, where $K\to1/2$, is obscured due to finite-correlation-length effects. We denote the points where $K=1/2$ with dots: blue dots give the points for which $n>1$, red dots give the points for $n<1$, and the black dot is an extrapolation to find the Mott lobe tip, where $n=1$. %Note that $K\to 1$ as one approaches the base of the Mott lobe in the superfluid phase. This is the hard-core regime, where the bosonic system can be mapped onto non-interacting fermions (with $K=1$) using a Jordan-Wigner transformation.
    }
    \label{fig:luttingerParam}
\end{figure}

The superfluid density can be used to extract the 
%itself is not sufficient to compute the 
Luttinger parameter: $\rho_s$ is proportional to the characteristic velocity of phase fluctuations, $v_j=u/K$ (see Secs.~\ref{sec:LL} and~\ref{sec:calcSF}).
To extract $K$, one needs to also calculate the charge compressibility 
$\kappa=\partial n/\partial\mu=1/\hbar\pi v_n$~\cite{giamarchi}, where $v_n=uK$ is the  
 characteristic velocity of density fluctuations.
 %, $v_n=uK$. This is related to the charge compressibility via $\kappa=\partial n/\partial\mu=1/\hbar\pi v_n$~\cite{giamarchi}. 
%We approximate the compressibility using a discrete derivative of the particle density at different chemical potentials. 
%We discuss the accuracy of this method and compare it to the value of $K$ extracted from $\langle a_ia^\dagger_j\rangle$ in Appendix~\ref{sec:paramFits}.
In Fig.~\ref{fig:luttingerParam}, we compute the Luttinger parameter across the zero-temperature phase diagram, approximating the compressibility using a discrete derivative of the density. The accuracy of this technique and associated error bars are discussed in Appendix~\ref{sec:paramFits}. In the superfluid phase, $K\leq 1$ due to the short-range nature of the Hubbard interactions~\cite{giamarchi,monienNN}. The Luttinger parameter approaches 1 at the density-driven Mott transition. By contrast, one expects $K\to 1/2$ at the Mott lobe tip, where the transition is BKT-like.
In Fig.~\ref{fig:luttingerParam}, we identify the contour along which $K=1/2$ with dots. Blue dots denote the points on the contour for which the particle density $n>1$ and red dots denote the points where $n<1$. We extrapolate to find the intersection of the $n>1$ branch with the contour of unit density ($n=1$) to approximate the position of the BKT transition. We find $t_c/U\approx 0.29$, which is in good agreement with previous numerical investigations~\cite{KRUTITSKY2016}. We expect $t_c$ to be pushed to larger values as the bond dimension is increased~\cite{reentrance}. As with other features of the BKT transition, scaling arguments are required to extract the precise location of the transition point.
%As with other features of the BKT transition, scaling is required to extract this behavior from the data in Fig.~\ref{fig:luttingerParam}.

At the base of the Mott lobe ($t,\mu\to 0$), the Luttinger parameter is well behaved while the superfluid density (as seen in Fig.~\ref{fig:heatMap}) rapidly changes as a function of $\mu/U$.  In particular, the limit $(t/U,\mu/U)\to(0,0)$ is singular, with the superfluid fraction taking on any value between $0$ and $1$ depending on the ratio $\mu/t$.
%how it is approached.
In the vicinity of the vacuum line ($\mu\to -2 t$), the density is small and the effects of the lattice can be ignored.  Thus, as expected for a translationally invariant system, the superfluid fraction approaches unity~\cite{leggettZeroTempSFDensity}.  Conversely, at the Mott transition  ($\mu\to 2t$ for small  $t/U$) the superfluid density vanishes. One can  interpret the point $(t/U,\mu/U)=(0,0)$ as the hard core limit, $U\to\infty$.  This  
lattice analog of the Tonks-Girardeau gas~\cite{girardeau1960} maps directly onto non-interacting fermions~\cite{jordanWigner}.
Figure~\ref{fig:luttingerParam} shows that $K\to 1$ in this limit, as one expects for non-interacting fermions.

%In the vicinity of the vacuum line ($\mu\to -2 t$), the density is small and the effects of the lattice can be ignored, so the superfluid fraction approaches unity. This can be seen in Fig.~\ref{fig:heatMap}. The limit $(t/U,\mu/U)\to(0,0)$ is singular because the superfluid density can take any value between 0 and 1 depending on the ratio $\mu/t\in(-2,2)$. In particular, 

Using the relationship between the zero-temperature superfluid density and the Drude weight (see Eq.~(\ref{eq:scalingFunc})), one finds that the superfluid density in the hard-core limit is given by~\cite{KRUTITSKY2016}
\begin{equation}
    \rho_s^{\rm HC}(\rho_0)=\frac{\sin(\pi\rho_0 d)}{\pi d}
\end{equation}
where $d$ is the lattice spacing and the particle density, $\rho_0$, is identical to that of a non-interacting fermions in 1D:
\begin{equation}
    \rho_0(\mu/t)=\frac{1}{\pi d}\arccos(-\mu/2t).
\end{equation}
%This formalizes the intuition that $\rho_s^{\rm HC}/\rho_0\to 1$ as one approaches the vacuum line ($\mu\to-2t$) while $\rho_s^{\rm HC}/\rho_0\to 0$ as one approaches the Mott lobe ($\mu\to2t$).

\subsection{\label{sec:entanglement}Entanglement}
As described in Sec.~\ref{sec:LL}, the Luttinger liquid phase of the 1D Bose-Hubbard model is a gapless critical phase. As such, the entanglement entropy between a region of length $L$ and the rest of the system scales as $S=(c/6)\log(L)$, where $c$ is the conformal charge~\cite{Calabrese_2004}. For a Luttinger liquid, $c=1$. 
%If the reduced density matrix of the length-$L$ subsystem is given by $\rho_L$, then the entanglement entropy is given by 
%\begin{equation}
%    S=-{\rm Tr}~\rho_L\ln\rho_L.
%\end{equation}
%This is, however, just one way of characterizing the entanglement of the system. A richer characterization is obtained from the full eigenvalue spectrum of $\rho_L$, which is known as the entanglement spectrum. A uMPS with bond dimension $\chi$ is a translationally-invariant wavefunction in which the reduced density matrix of a bipartition of the system has exactly $\chi$ eigenvalues. Enforcing that $\chi$ is finite reduces the computational complexity of the problem, but it also constrains $S\leq\ln(\chi)$. Thus, the uMPS with finite $\chi$ can never capture the entanglement of an infinite system in the Luttinger liquid phase.
In the thermodynamic limit the entanglement entropy should  diverge; at finite bond dimension ($\chi$), however, our matrix product state has a finite correlation length ($\xi$) that cuts off the entanglement. One therefore expects that for large $\xi$~\cite{CalabreseLefevre,jMooreEE},
%While truncation of the entanglement spectrum is the natural interpretation of the MPS ansatz, one can also interpret the finite bond dimension $\chi$ as imposing a finite correlation length. This can be seen naively by comparing the strict upper bound on the entropy with the conformal scaling relation. Given the entanglement entropy of a converged uMPS with bond dimension $\chi$, one can define a correlation length, $\tilde\xi$:
\begin{eqnarray}
    S(\chi)&=&(c/6)\ln(\xi(\chi))\\
    &\sim& (\kappa c/6)\ln(\chi),\label{eq:S_xi}
    %\implies\tilde\xi(\chi)\leq \chi^{6/c}.
\end{eqnarray}
%This correlation length is only defined up to an overall prefactor~\cite{Calabrese_2004}, so we use the tilde to differentiate it from Eq.~(\ref{eq:xi}).
%The correlation length $\tilde\xi$ obeys a strict upper bound, $\tilde\xi(\chi)\leq \chi^{6/c}$.
% If one defines the effective length scale as the correlation length, $\tilde\xi$, it is constrained by the bond dimension:
% \begin{equation}
%     S(\chi)=(c/6)\ln(\tilde\xi(\chi))\implies\tilde\xi(\chi)\leq \chi^{6/c}.
% \end{equation}
% This correlation length $\tilde\xi$ is only defined up to an overall prefactor~\cite{Calabrese_2004}, so we use the tilde to differentiate it from Eq.~(\ref{eq:xi}). 
where  we  have used the relation
$\xi\sim\chi^\kappa$ with
\begin{equation}
\kappa=\frac{6}{c (\sqrt{12/c}+1)}.
\label{eq:kappa}
\end{equation}
%For systems with a conformal symmetry, Calabrese and Lefevrer argued that the eigenvalues of the reduced density matrix satisfy a scaling argument that depends only on their conformal charge~\cite{CalabreseLefevre}. Pollmann \textit{et al.} used this result to approximate the excitation gap of the uMPS and determine the scaling of the correlation length with bond dimension~\cite{jMooreEE}. They find that
%\begin{equation}
%    \tilde\xi(\chi)\propto\chi^\kappa\hspace{1cm}\kappa=\frac{6}{c(\sqrt{12/c}+1)}
%    \label{eq:kappaDefn}
%\end{equation}
%for systems with a conformal charge $c$.

%In a gapless critical system, which has an infinite correlation length, one can define the MPS correlation length in terms of the entropy of the optimized MPS with bond dimension $\chi$: $S(\chi)=(c/6)\ln(\tilde\xi(\chi))$. Note that $\tilde\xi$ is defined up to a non-universal prefactor, given that the conformal formula for the entropy has non-universal additive corrections~\cite{Calabrese_2004}. Nonetheless, this identification is sufficient to derive the scaling of $\xi$ with bond dimension. The MPS constraint implies that the correlation length has an upper bound, $\tilde\xi(\chi)\leq \chi^{6/c}$. A more careful calculation in Ref.~\cite{jMooreEE}, based on a conformal scaling argument proposed in Ref.~\cite{CalabreseLefevre}, shows that $\tilde\xi(\chi)\propto \chi^\kappa$ for large $\chi$ where the exponent $\kappa=6/(\sqrt{12c}+c)$.

\begin{figure}
    \centering
    \includegraphics[width=3.375in]{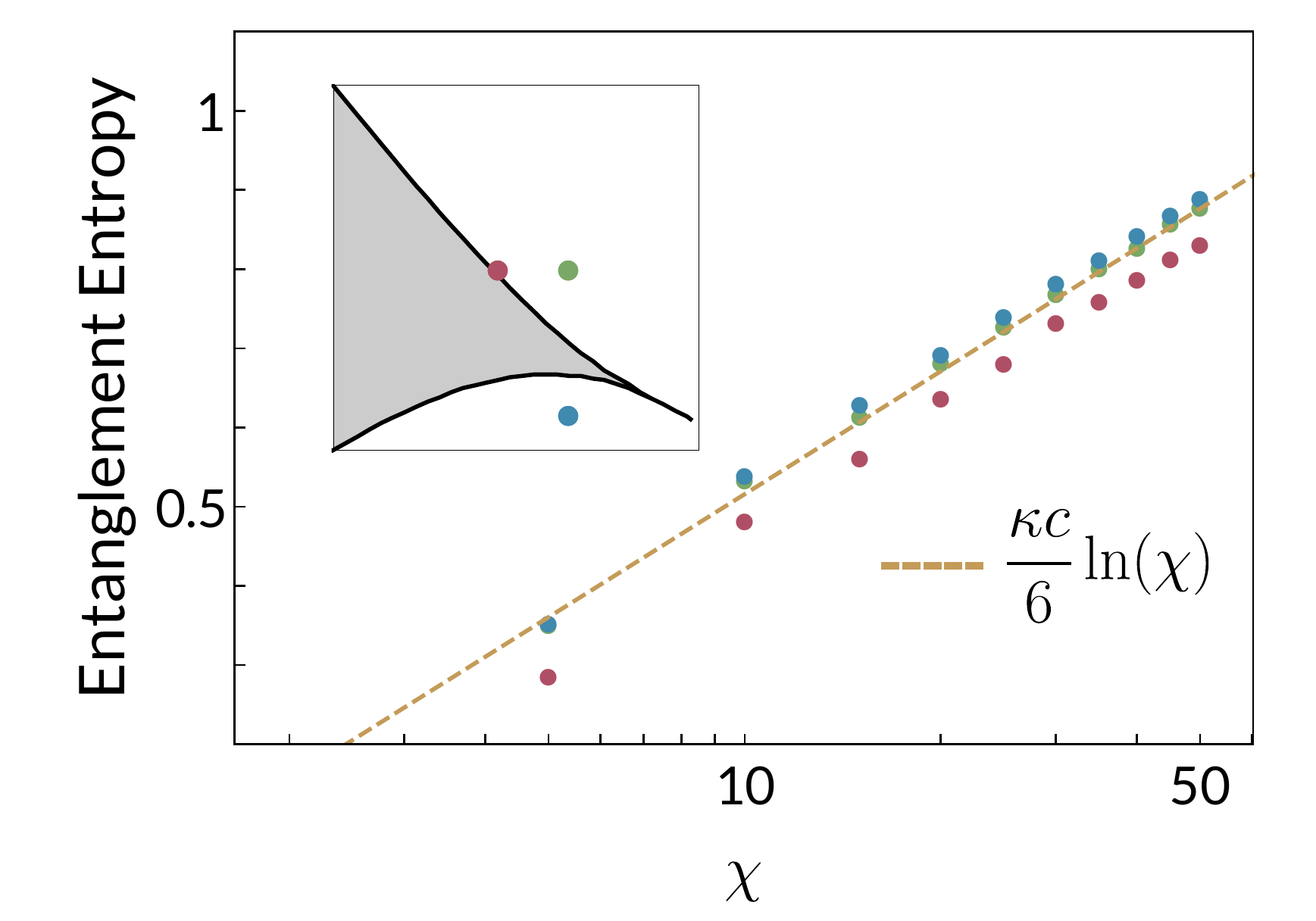}
    \caption{Plot of the entanglement entropy between bipartitions of the infinite system versus bond dimension on a semi-log scale. Colored dots in the inset show the points on the $t/U-\mu/U$ phase diagram where data was taken.  Horizontal  and vertical ranges  of the inset's axes roughly correspond to those of Figs~\ref{fig:heatMap} and~\ref{fig:luttingerParam}. We observe the expected logarithmic scaling of the entropy with bond dimension. The dashed line shows the expected scaling of the entanglement entropy based on the calculation in Ref.~\cite{jMooreEE}. We find excellent agreement with this prediction.
    }
    \label{fig:EEFig}
\end{figure}

In Figure~\ref{fig:EEFig}, we plot the entanglement entropy versus bond dimension on a semi-log plot at a few representative points in the Luttinger liquid phase. We find good agreement with the prediction in Eqs.~(\ref{eq:S_xi}) and~(\ref{eq:kappa}). Furthermore, the scaling of the correlation length agrees with that shown in the inset of Fig.~\ref{fig:densityMatFig}, which is determined entirely from the correlation function $\langle a_ia^\dagger_j\rangle$.

%There are, therefore, two characteristic parameters that determine the scaling properties of the 1D Bose-Hubbard model: the Luttinger parameter, $K$, which controls the behavior of correlation functions; and the conformal charge, $c$, which controls finite-length and finite-entanglement scaling of the system. While $c$ (and $\kappa$) are universal parameters in a Luttinger liquid, the Luttinger parameter can vary continuously throughout the phase diagram. A few parameters exhibit scaling that depends on both $\kappa$ and $K$. We discuss one such example, the quasicondensate density, in Appendix~\ref{sec:quasicondensate}. 

\section{\label{sec:exp}Experimental Applications}
As described in Sec.~\ref{sec:LL}, the superfluid density is a natural observable in bulk superfluid helium.  Measuring the superfluid density in a 1D lattice system, however, is  more challenging.  The two most promising settings are helium or ultracold atoms.  In fact, one dimensional superfluidity has been studied using 
%Measuring the superfluid density of one-dimensional systems has historically been a challenge. There are at least two arenas where this is being done. The first is Helium nanopores, where
%. Experiments involve adsorbing 
$^4{\rm He}$ adsorbed in %hexagonal 
channels a few nanometers in diameter~\cite{heSFExpt2}. The 1D regime is reached when the thermal wavelength of the $^4{\rm He}$ is large compared to the diameter of the tubes. This condition can be understood as the freezing out of transverse modes, which are gapped due to their quantization. The array of 1D tubes is then placed on a torsional oscillator with a sufficiently low frequency of oscillation to only excite longitudinal modes of the nanotubes~\cite{heSFExpt1,heSFExpt2,heSFExpt3}. One can then extract a frequency shift that is directly related to the superfluid density.
%The results of these experiments were discussed briefly in Sec.~\ref{sec:neq}.  
Unfortunately, in this setting there is no simple way to add a lattice or control the interaction strength.

%The benefit of these experiments is that they are easily interpretable, owing in part to their similarity to seminal 3D experiments by Andronikashvili~\cite{andro,andro2,androSS}. What they gain in interpretability, however, they lack in flexibility -- use of $^4{\rm He}$ means that one has no control over the relative interaction strength.
%nor the ability to realize the Bose-Hubbard (tight-binding) regime. 
%As there is no lattice in these experiments, they are not capable of testing the predictions made here.

Cold atoms
%the second of the two arenas, have a distinct advantage in this regard. 
can overcome both of these issues.
The 1D Bose-Hubbard Hamiltonian is the natural description of bosonic atoms in a deep optical lattice. Moreover, the contact interactions can be tuned by a Feshbach resonance~\cite{expReview1,expReview2}. Again, the 1D regime is realized by applying sufficiently large transverse confinement, freezing out the transverse modes. Unfortunately,  measuring the superfluid density in cold atom systems is difficult. 
%The techniques historically used for liquid helium do not  apply.

%Paradoxically, although a finite superfluid density was historically the primary way to identify the superfluid phase, cold atom systems tend to do so by measuring the (often-associated) condensate density. This is because measuring $\langle n_k\rangle$ is straightforward by a mapping of the momentum space density onto the position space density, known as time-of-flight imaging~\cite{expReview1}. Applying rotations to cold atom systems, as in the torsional oscillator experiments, is quite finnicky by comparison. Furthermore, as atoms are trapped in optical lattices, it is not immediately clear how one would measure their moment of inertia.

In the absence of a lattice, there have been at  least four ways  to meet this challenge: (1) Collective modes; (2) Density response to rotation; (3) Spectral response to an artificial vector potential; and (4) Velocities of first and second sound.  We briefly describe each of these. So far all experiments and proposals have been in either 3D or 2D.

Adapting these approaches to a 1D Bose-Hubbard system would require substantial work:
Method (1) does not have an obvious analog in 1D.
Methods (2)-(3) 
%Applying (1)-(3) to the one-dimensional lattice introduces a variety of practical challenges. 
%For example,  all of these
%any steady-state  probe of the 1-D superfluidity 
would require a periodic ring geometry~\cite{zoller2016}, which has not been realized with a lattice.  
%While there have been proposals involving such geometries \cite{}, they have not yet been realized.
%It is therefore most likely that probes of superfluidity in finite-sized 1D systems will take place out of the steady-state regime. %While analogs of ``stirring" (``kicking") and artificial vector potentials (Peierls' phases) exist 1D systems, the inability to drive persistent currents means that any analysis would inevitably have to take the trap into account.
Method (4) does not apply in 1D.
%would require having a model of how the sound modes in a 1D lattice relate to the superfluid density.  
We briefly elaborate on each of these.
%, noting that there currently exists no simple approach to measuring the superfluid density of a 1D lattice system.

{\em (1) Collective modes:}
%In higher dimensions, t
The lowest energy %collective 
mode of a gas in an anisotropic trap is analogous to the fundamental mode of a torsional oscillator, and hence provides information about superfluidity~\cite{stringari1,stringari2,stringari3,stringari4,stringari6,Li2012,stringari5,Perrin}. For example, the precession frequency of quadrupole modes has been used to extract the moment of inertia of the unitary Fermi gas~\cite{GrimmExpt1} and of dipolar bosons~\cite{modugno} in three dimensions.
%the Grimm group has been able to use this technique to extract the superfluid density of the 3D unitary Fermi gas~\cite{GrimmExpt1}.
The superfluid fraction can then be determined by the reduction of the moment of inertia from that of a rigid body.
%They stirred the gas using a deformation in the trapping potential, thereby inducing rotation, then measured the moment of inertia using the precession frequency of a hydrodynamic quadrupole mode. 
This technique relies on a hydrodynamic description of the cloud, and hence requires sufficiently strong interactions.
%g gas is preferable in its ability to reach this regime on shorter timescales. 
One major challenge here is that the trapped system is highly inhomogeneous, and the measured superfluid fraction is spatially averaged.  Driving the collective modes can also
%The method of measurement is liable to induce 
heat the sample or excite vortices.
%This technique relies on the gas having strong interaction, and the stirring and measurement techniques are liable to induce heating and excite vortices. %{\color{red} Does this work in lattice?}
Further, 
this technique is not directly applicable in the presence of an optical lattice.  The lattice breaks rotational symmetry, which complicates the relationship between 
%using the frequency dependence of 
the frequencies of the quadupole modes and the superfluid density.
%to determine the superfluid density relies on cylindrical symmetry.  An optical lattice that would be broken by the presence of an optical lattice.

{\em (2) Density response to rotation:}
Ho and Zhou argued that the local superfluid density in three dimensions can be extracted from the response of the column density profile to rotation~\cite{hoZhou}. 
Importantly, their approach directly  
%With that said, the imaging 
%techniques which 
gives the spatial dependence of the superfluid density in an inhomogeneous trap.
The derivation, however, relies on a strictly harmonic trapping potential and would need to be modified to include a lattice. 
%Ho and Zhou proposed a method of measuring the superfluid density in a 3D gas by measuring the response of the columnated density profile to rotation~\cite{hoZhou}. This technique for measuring superfluid density is useful as there are reliable ways to image the particle density with minimal disturbances.

{\em (3) Spectral response to an artificial vector potential:}
Rather than stirring a trapped gas with a potential deformation \cite{GrimmExpt1,modugno}, one can probe superfluidity by introducing an Raman-induced artificial vector potential~\cite{spielmanProposal,spielmanGauagePotentialExpt}:
%as was done in the Refs.~\cite{GrimmExpt1} and~\cite{modugno}, one can also induce rotation through the introduction of an artificial vector potential. Raman driving schemes~\cite{spielmanProposal,spielmanGauagePotentialExpt} have been proposed and realized as means to introduce an effective vector potential into the Hamiltonian. 
A set of Raman lasers dresses the atomic states in such a way that they experience an artificial magnetic field. 
Cooper and Hadzibabic~\cite{zoranCooper1,zoranCooper2} %subsequently proposed an application of this technique to measure the superfluid density spectroscopically. They 
showed that the superfluid density can be determined from the populations of the Raman-dressed bands.  This enables a spectroscopic determination of superfluid density which can potentially be spatially resolved~\cite{Lin2018}.  

{\em (4) Velocities of first and second sound:}
Recent experiments by the Grimm~\cite{GrimmSS} and Hadzibabic~\cite{zoranSS} groups have determined the superfluid density in 2D systems by measuring the velocities of first and second sound. Similar sound-speed measurements 
%are well 
%Such procedures appear reasonably robust and naturally 
%suited to the (non-equilibrium) strengths of cold atom experiments, and 
can be carried out in
%directly applicable to 
lattice gases. 
%The primary drawback of this method is that it requires %heavily 
%extensive knowledge of the physical system -- the sound velocities can only be related to the superfluid density if one knows the temperature, specific heat, entropy, and the isothermal and isobaric compressibilities. Refs.~\cite{GrimmSS} and~\cite{zoranSS} make use of previously-derived relationships in order to determine these parameters with a minimal number of measurements. In the absence of such relations, the procedure becomes quite cumbersome.
%While Method (4) is most naturally applicable to the 1D gas from an experimental point of view, 
%An important complication is that
Unfortunately,
the two-fluid hydrodynamics of a Luttinger liquid differs from that of higher-dimensional superfluids~\cite{secondSoundLL1,secondSoundLL2,secondSoundLL3,secondSoundLL4}. Specifically, in dimension $d$, as $T\to 0$ 
the ratio of the velocities
%it is predicted in dimensions $d>1$ that
first and second sound modes 
%decouple and their ratio 
approach $\sqrt{d}$. As these modes travel at the same velocity in one dimension, they do not fully decouple,
invalidating the analysis that was used to find the superfluid density of the 2D systems.
%at zero temperature 
%and the observed sound waves are hybrids of first and second sound. 
%Thus, it is unclear how to modify this procedure to measure the superfluid density in one-dimensional systems.

\section{\label{sec:conclusions}Summary}
We have provided a comprehensive discussion of superfluidity in the 1D Bose-Hubbard model, aided by numerical simulations with infinite matrix product state techniques. %We argued that t
The zero-temperature superfluid fraction is related to a Drude weight, which we measure directly from the response to a phase twist. We give some discussion of both finite temperature and finite size considerations, and how they depend on dimension.
%This parameter fixes the equilibrium superfluid density as a function of length and temperature (see Eq.~(\ref{eq:scalingFunc})) and is also pertinent for alternative dynamical definitions of the superfluid density. %In this way, unlike in higher dimensions, the superfluid density in one dimension is directly related to the long-distance behavior of the single-particle density matrix.

Our work demonstrates the success of using infinite matrix product state techniques to model gapless critical systems. We illustrate a specific advantage of VUMPS over iDMRG in such systems, namely the ability to efficiently capture long-range correlations and entanglement even after a small number of iterations.

In addition to calculating superfluid densities, we use several independent approaches to extract the Luttinger parameters which parameterize all long-wavelength properties of the gas.  These disparate approaches show non-trivial behavior and agree with one-another.  Furthermore, we explore connections between finite size scaling and finite entanglement scaling.

%We explore the behavior of the single particle density matrix, extracting a Luttinger parameter from the spatial dependence.  We extract the other Luttinger parameter from the chemical potential dependence of the density.

%Furthermore, 
%although an MPS with finite bond dimension cannot capture certain properties of the true ground state, 
%we extract the Luttinger parameters from our ansatze, which describe all long-wavelength properties.
%and thus infer all properties of the model that are quantitatively missed by the tensor network approximation.
%We assess the reliability of the Luttinger parameters through finite-entanglement scaling.
%, which shares many of the properties of finite-length scaling with an effective correlation length $\xi\sim\chi^\kappa$~\cite{jMooreEE}.

\begin{acknowledgments}
We thank Jim Sethna, Joel Moore and Matt Fishman for helpful conversations. This material is based upon work supported by the National Science Foundation under Grant No. PHY-2110250.
\end{acknowledgments}

\appendix

\section{\label{sec:LLHelicityModulus}Helicity modulus of a Luttinger liquid}
%We include a derivation of the helicity modulus of a Luttinger liquid, as reported in Ref.~\cite{affleckQMC}. 
The helicity modulus, $\Upsilon(L,T)$, of a 1D system of length $L$ and temperature $T$ is defined as
\begin{equation}
    \frac{F[L,T,\Phi]-F_0[L,T]}{L}=\frac{1}{2}\Upsilon(L,T)\left(\frac{\Phi}{L}\right)^2+\cdots
    %\mathcal{O}\left(\frac{\Phi}{L}\right)^4
    \label{eq:upsilon1d}
\end{equation}
where $F$ is the free energy and $\Phi$ is the phase twist across the periodic boundaries, $\Psi(x+L)=e^{i \Phi}\Psi(x)$. The omitted terms scale as $(\Phi/L)^4$. As reported in \cite{affleckQMC}, this quantity can be exactly calculated for a Luttinger liquid, described by a Hamiltonian
\begin{equation*}
    \frac{\mathcal{H}_{LL}}{\hbar}=\sum_{q\neq 0}\omega_qb^\dagger_qb_q+\left(\frac{\pi}{2L}\right)\left(v_jJ^2+v_n(N-N_0)^2\right).
\end{equation*}
%so the free energy can be evaluated explicitly. 
As described  in Section~\ref{sec:LL}, $J$ is the winding number, $N$ the number of bosons, and $b_q$ are excitations of momentum $q$.
Here we present an explicit derivation of the resulting 
%detailed derivation of the 
helicity modulus.
%for this model.

We begin by noting that the partition function $Z=\exp(-\beta F)$ factors into the product $Z_bZ_JZ_N$, corresponding to contributions from each term in the Hamiltonian. Of these, only the topological phase twist term will be affected by the boundary condition twist. The twist is incorporated by requiring  $J=2 j-\Phi/\pi$ where $j$ is an integer, giving us
%\begin{equation}
%    Z_J=\sum_{j=-\infty}^\infty \exp\bigg(-\beta\frac{2\pi v_j}{L}j^2\bigg).
%    \label{eq:zj0}
%\end{equation}
% We  therefore only need to consider this term.
%
%The winding number $J\in 2\mathbb{Z}$ for periodic boundary conditions, such that the wavefunction obeys $\Psi(x+L)=e^{i\pi J}\Psi(x)$. Adding a boundary condition twist $\Phi$ means that $\Psi(x+L)=e^{i\pi J+i\Phi}\Psi(x)$. This can be accommodated by taking $J\to J+\Phi/\pi$ in Eq.~(\ref{eq:zj0}):
\begin{eqnarray}
    Z_J(\Phi)&=&\sum_{j=-\infty}^\infty \exp\bigg(-\beta\frac{2\pi v_j}{L}(j+\Phi/2\pi)^2\bigg)\\
    &=&\sqrt{\frac{LT}{2\pi\Upsilon_0}}\vartheta_3(\Phi/2,e^{-LT/2\Upsilon_0})
\end{eqnarray}
where $\vartheta_3(z,q)=\sum_{n=-\infty}^\infty q^{n^2}e^{2niz}$ is the Jacobi theta function of the third kind and  
%It is also useful to note that
%Note that the partition function is a periodic function of $\Phi$, as must be the case. These sums can be written as
%\begin{equation}
%    \begin{split}
%        Z_J(0)&=\vartheta_3(0,e^{-2\pi^2\Upsilon_0/LT})\\
%        &=\sqrt{\frac{LT}{2\pi\Upsilon_0}}\vartheta_3(0,e^{-LT/2\Upsilon_0}).
%        \label{eq:zj0jacobi}
%    \end{split}
%\end{equation}
%\begin{equation}
%    Z_J(\Phi)=\sqrt{\frac{LT}{2\pi\Upsilon_0}}\vartheta_3(\Phi/2,e^{-LT/2\Upsilon_0})
%\end{equation}
%where they are represented in terms of Jacobi theta function of the third kind,
%\begin{equation}
%    \vartheta_3(z,q)=\sum_{n=-\infty}^\infty q^{n^2}e^{2niz}.
%\end{equation}
%Note the change of variables to
$\Upsilon_0=\hbar v_j/\pi$ is the zero-temperature helicity modulus. 
%The two expressions in Eq.~(\ref{eq:zj0jacobi}) are related by completing the square. 

We now Taylor expand the ratio of theta functions for small twist angles, finding
\begin{equation}
    \ln\bigg(\frac{Z_J(\Phi)}{Z_J(0)}\bigg)=\frac{1}{8}\frac{\vartheta_3^{\prime\prime}(0,e^{-LT/2\Upsilon_0})}{\vartheta_3(0,e^{-LT/2\Upsilon_0})}\Phi^2+\mathcal{O}(\Phi)^4
    \label{eq:taylorLogZ}
\end{equation}
where $\vartheta_3^{\prime\prime}(z,q)=\partial^2_z\vartheta_3(z,q)$. Finally, by substituting Eq.~(\ref{eq:taylorLogZ}) back into Eq.~(\ref{eq:upsilon1d}), we obtain an expression for the helicity modulus:
\begin{equation}
    \begin{split}
        \Upsilon(L,T)/\Upsilon_0&=-\frac{LT}{4\Upsilon_0}\frac{\vartheta_3^{\prime\prime}(0,e^{-LT/2\Upsilon_0})}{\vartheta_3(0,e^{-LT/2\Upsilon_0})}\\
        &=1+\frac{\pi^2\Upsilon_0}{LT}\frac{\vartheta_3^{\prime\prime}(0,e^{-2\pi^2\Upsilon_0/LT})}{\vartheta_3(0,e^{-2\pi^2\Upsilon_0/LT})}.
    \end{split}
    \label{eq:upsilonScaling}
\end{equation}
The normalized helicity modulus $\Upsilon/\Upsilon_0$ is a scaling function that depends only on the quantity $LT/\Upsilon_0$. The two forms shown in Eq.~(\ref{eq:upsilonScaling}), both of which appear in the literature, are related by completing the square. The physical consequences of this result are discussed in Sec.~\ref{sec:limits}.

\section{VUMPS implementation \label{sec:vumps}}
Here we discuss our implementation of the VUMPS algorithm. We refer the reader to Refs.~\cite{vumps} and~\cite{vumps2} for further details and justification. We will follow the standard graphical notation for tensor networks~\cite{SCHOLLWOCKreview}. Throughout this section, graphical equations will show a finite portion of (what should be assumed to be) an infinitely long MPS.

In a given iteration, we begin with a uniform matrix product state. Rather than parameterizing the MPS in the uniform gauge, 
\begin{equation}
    \vcenter{\hbox{\includegraphics[scale = 0.33]{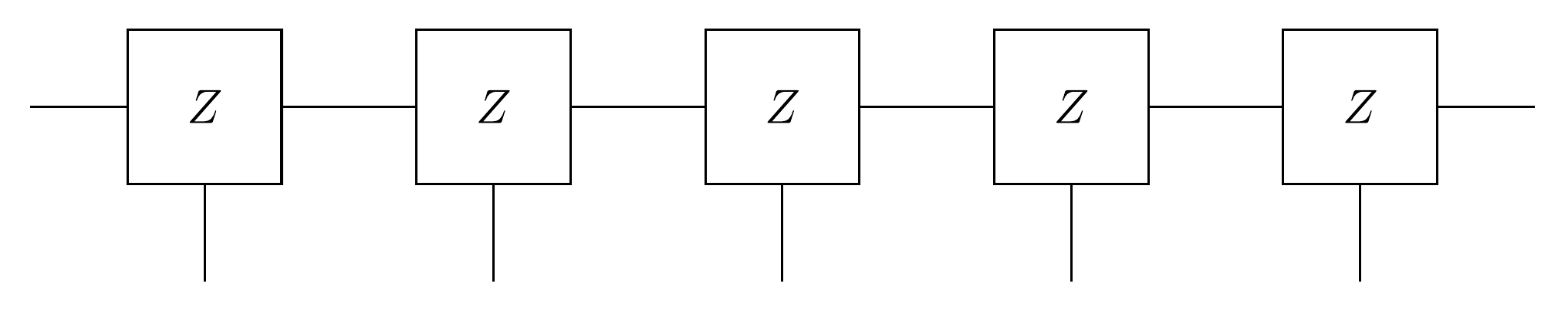}}},
    \label{eq:uniform}
\end{equation}
where the same tensor acts on each site, it is convenient to write the state in mixed-canonical form:
\begin{equation}
    \vcenter{\hbox{\includegraphics[scale = 0.33]{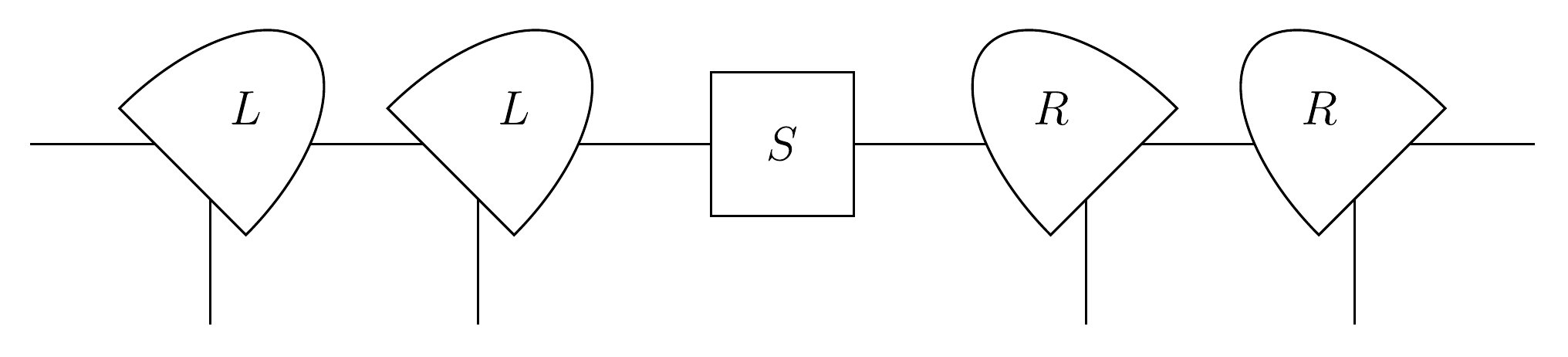}}}.
    \label{eq:mixedCanonical}
\end{equation}
The uniform and mixed-canonical forms are related by a gauge transformation~\cite{SCHOLLWOCKreview}. The mixed-canonical form is defined by three tensors, ${\bf L}$, ${\bf S}$, and ${\bf R}$. As indicated by the shape of their symbols, the tensors ${\bf L}$ and ${\bf R}$ are left and right-orthogonal tensors, obeying 
%. These conditions can be written graphically as
\begin{align}
    \vcenter{\hbox{\includegraphics[scale = 0.33]{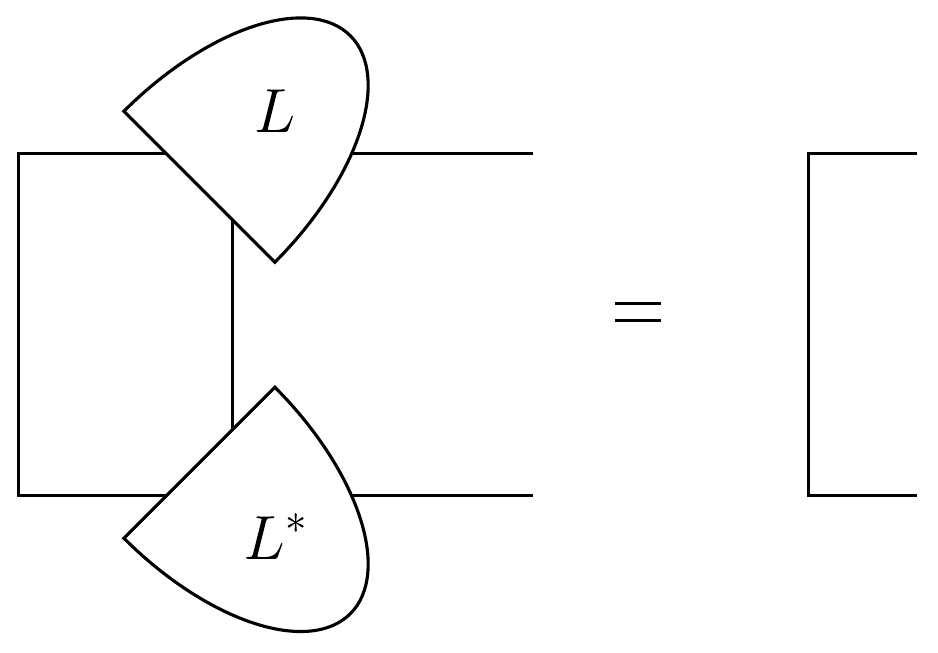}}}
    \label{eq:l_orth}\\
    \vcenter{\hbox{\includegraphics[scale = 0.33]{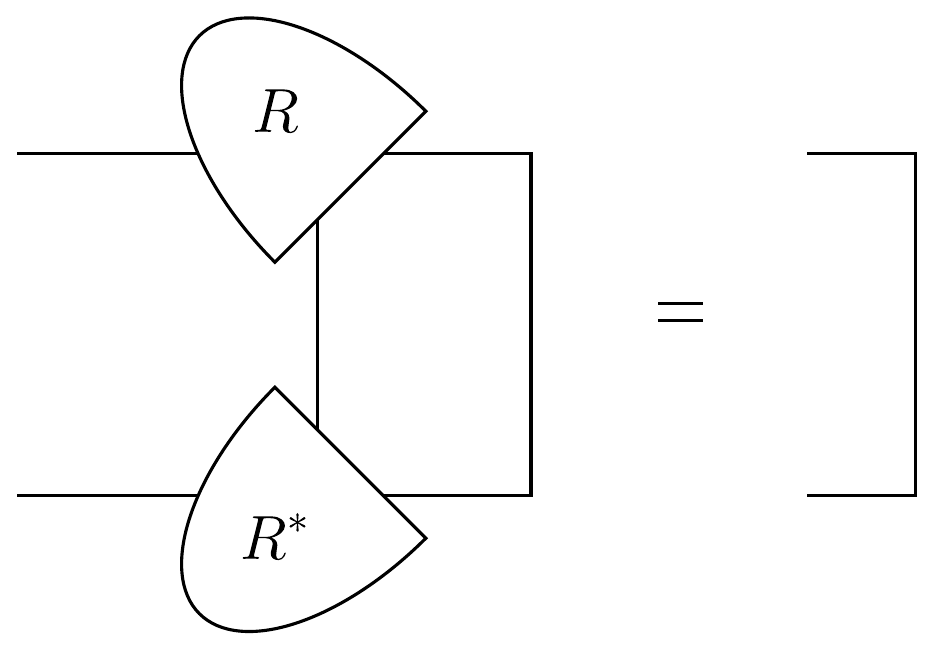}}}
    \label{eq:r_orth},
\end{align}
where the symbols on the right hand side represent identity tensors.
This orthogonality dramatically simplifies the calculation of expectation values and hence is the preferred way of storing and manipulating a matrix product state. In order for the state to be translationally-invariant, these tensors should satisfy
\begin{equation}
    \vcenter{\hbox{\includegraphics[scale = 0.28]{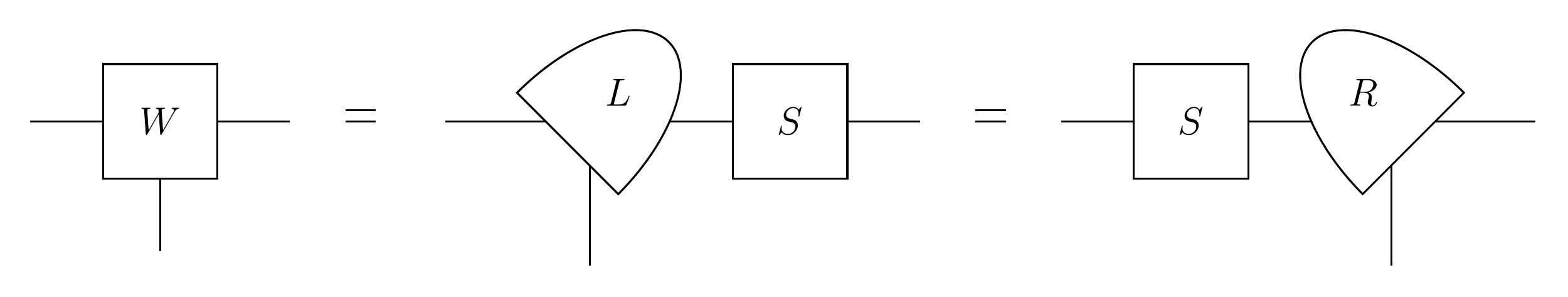}}},
    \label{eq:ti_equality}
\end{equation}
which defines the tensor ${\bf W}$.  The mixed-canonical form can also be written as
\begin{equation}
    \vcenter{\hbox{\includegraphics[scale = 0.33]{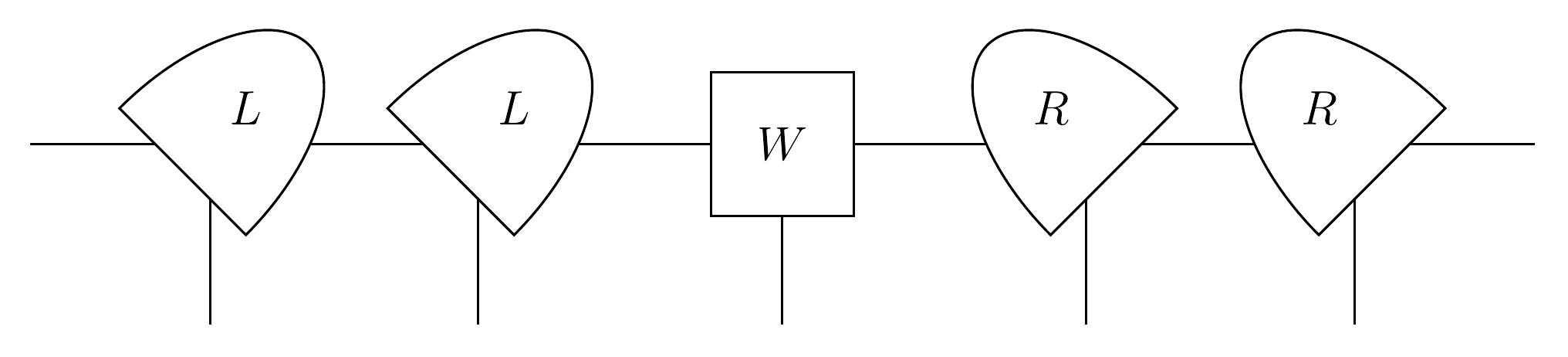}}}.
    \label{eq:mixedCanonicalW}
\end{equation}
%, defined by Eq.~(\ref{eq:ti_equality}), could just as well replace ${\bf S}$ in the definition of mixed-canonical form. 
Graphically, Eq.~(\ref{eq:ti_equality}) implies that one can freely shift the inversion center of the uniform MPS without changing any observable properties of the state:
\begin{equation}
    \vcenter{\hbox{\includegraphics[scale = 0.3]{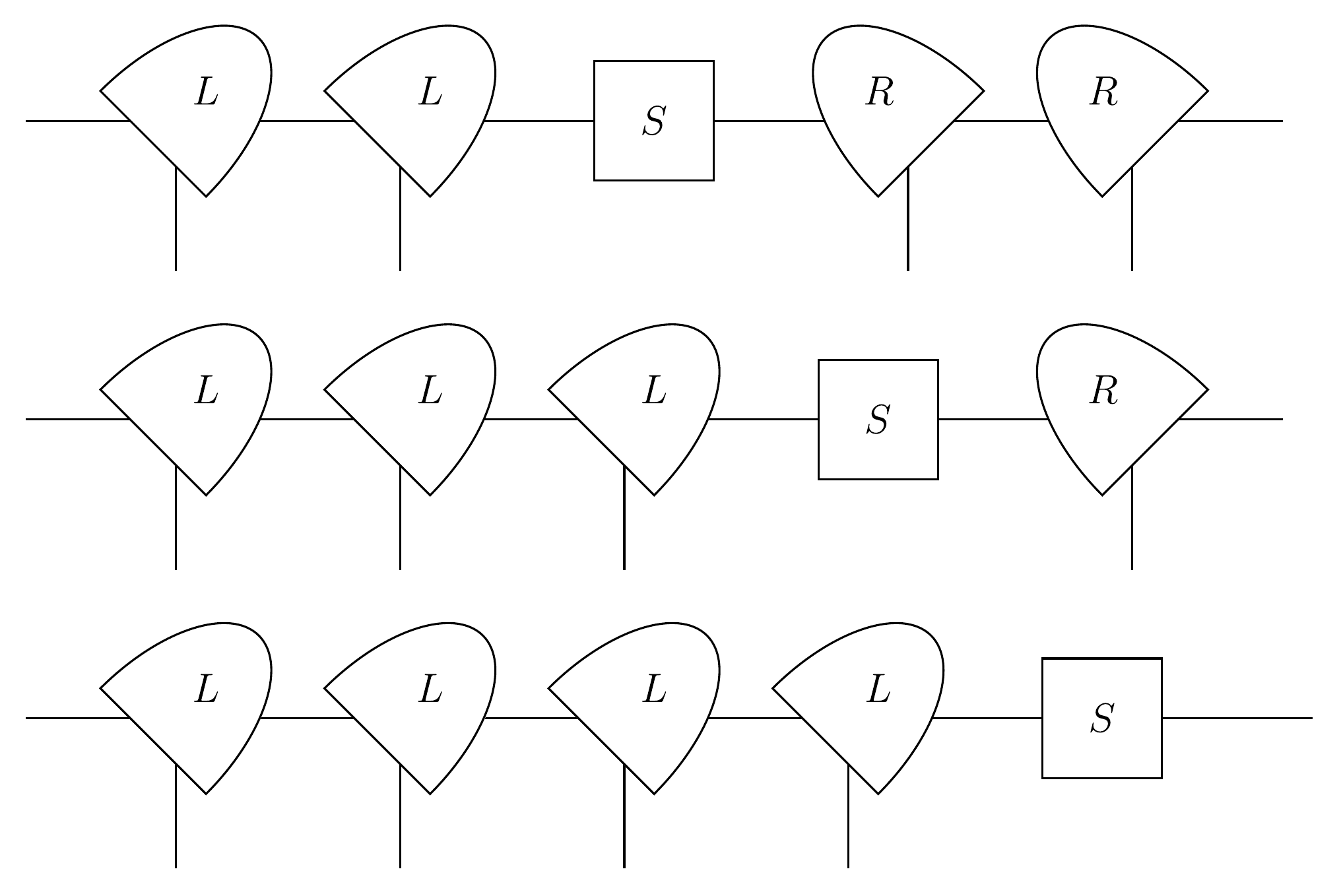}}}.
    \label{eq:imps}
\end{equation}
All three of the above states represent the same physical wavefunction.

A uniform matrix product state is defined by the set of tensors ${\bf L}$, ${\bf R}$, ${\bf W}$, and ${\bf S}$, obeying the constraint in Eq.~(\ref{eq:ti_equality}).  The VUMPS algorithm involves using energetic arguments to update ${\bf W}$ and ${\bf S}$, and linear algebra techniques to update ${\bf L}$ and ${\bf R}$.  It converges to a uniform matrix product state, but as schematically shown in Fig.~\ref{fig:schematic}, at intermediate stages the central site differs from the others.  We will discuss one step of the algorithm, going from 
$\{ {\bf L},{\bf R},{\bf W},{\bf S}\}$ to
$\{ {\bf \tilde L},{\bf \tilde R},{\bf \tilde W},{\bf \tilde S}\}$.

We find ${\bf \tilde W}$ and ${\bf \tilde S}$ by minimizing the energies,
\begin{eqnarray}
    {\cal E}_W&=&\vcenter{\hbox{\includegraphics[scale = 0.28]{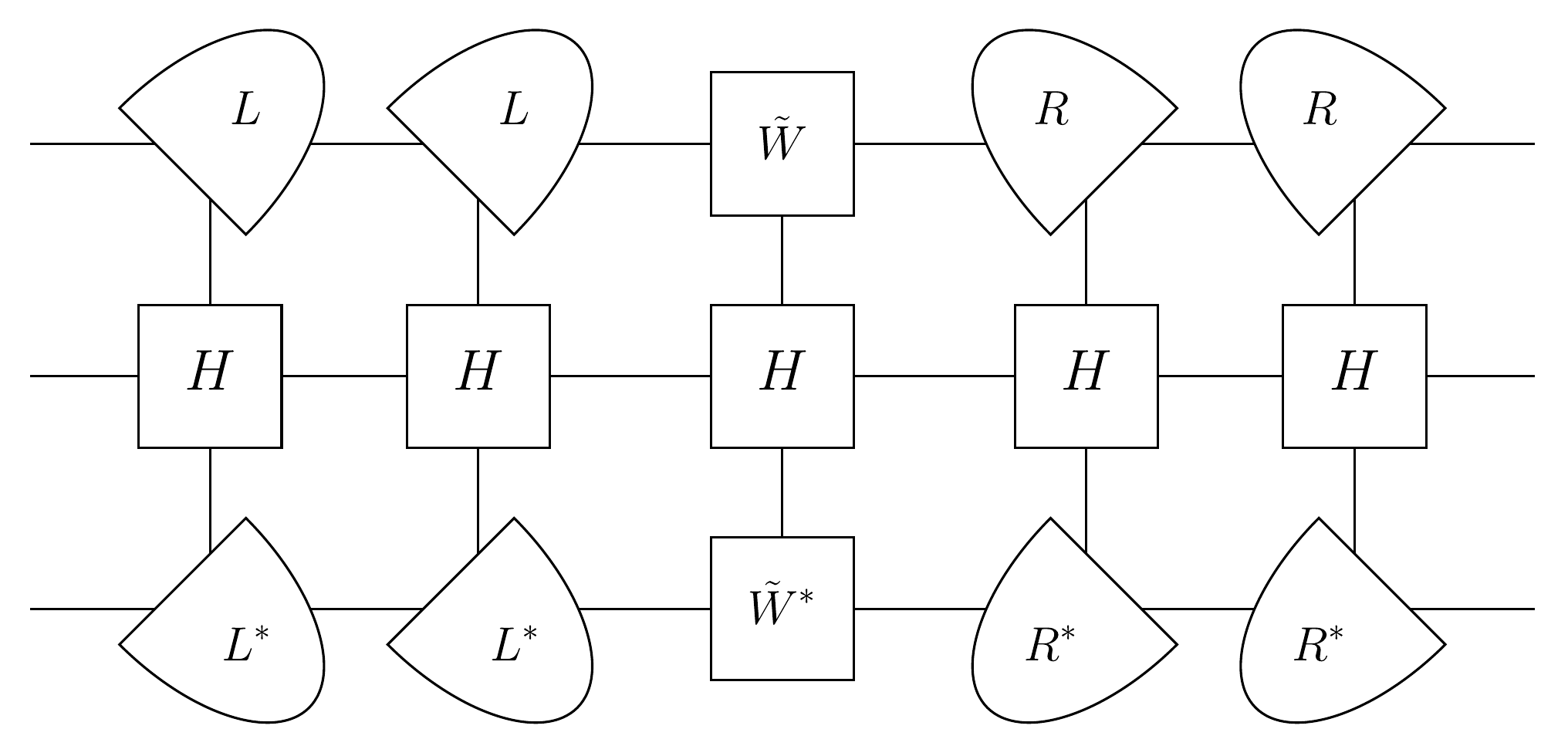}}},
    \label{eq:make_heff}\\
    {\cal E}_S&=&\vcenter{\hbox{\includegraphics[scale = 0.28]{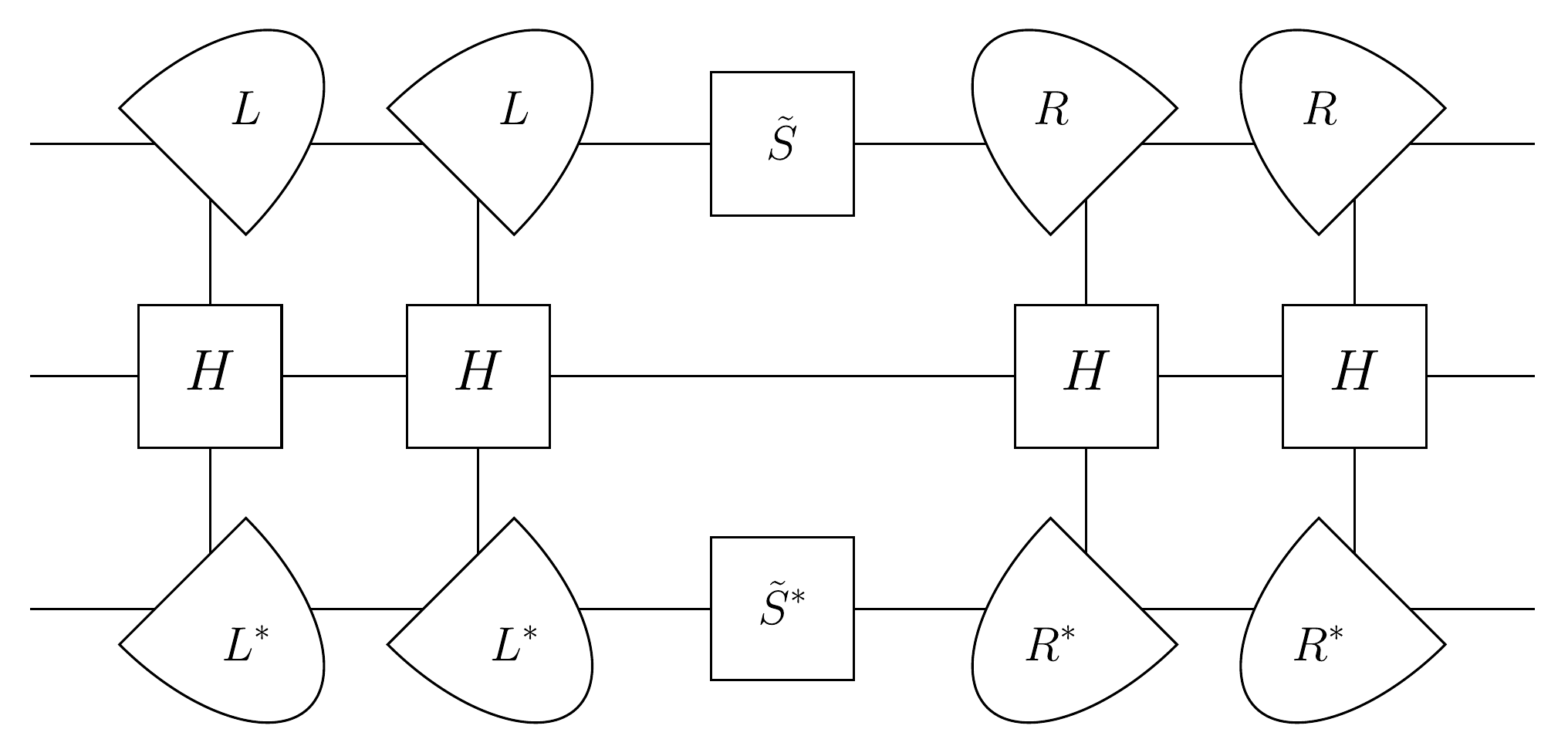}}}.
    \label{eq:make_heffS}
\end{eqnarray}
which correspond to expectation values of the Hamiltonian. In Eqs.~(\ref{eq:make_heff}) and~(\ref{eq:make_heffS}), the Hamiltonian is written as a matrix product operator (MPO)~\cite{SCHOLLWOCKreview}: 
%
%Given the tensors ${\bf L}$ and  ${\bf R}$ from a previous iteration, we find the new tensor ${\bf \tilde W}$
%
%
%compute the action of the Hamiltonian operator on this state in order to find the optimal next ``step" to take. Specifically, we consider the expectation value of the Hamiltonian with respect to an inhomogeneous state where only a single tensor differs from the rest. We'll call this tensor ${\bf \tilde W}$ and the inhomogeneous state $|\Psi_{\tilde W}\rangle$. We then compute the expectation value of the Hamiltonian, $\langle \Psi_{\tilde W}|\hat{H}|\Psi_{\tilde W}\rangle$, which in mixed-canonical form is given by
%\begin{equation}
%    \vcenter{\hbox{\includegraphics[scale = 0.33]{makeHeff.pdf}}}.
%    \label{eq:make_heff}
%\end{equation}
%The ${\bf H}$ tensors are matrix product operators (MPOs) that, when contracted, yield all the terms in the Hamiltonian acting on the MPS~\cite{SCHOLLWOCKreview}. %There are multiple allowed forms of an MPO representation of a given Hamiltonian. 
%The infinite MPO corresponding to Eq.~(\ref{eq:Hbh}) can be written as a matrix of matrices,
\begin{multline}
    \vcenter{\hbox{\includegraphics[scale = 0.33]{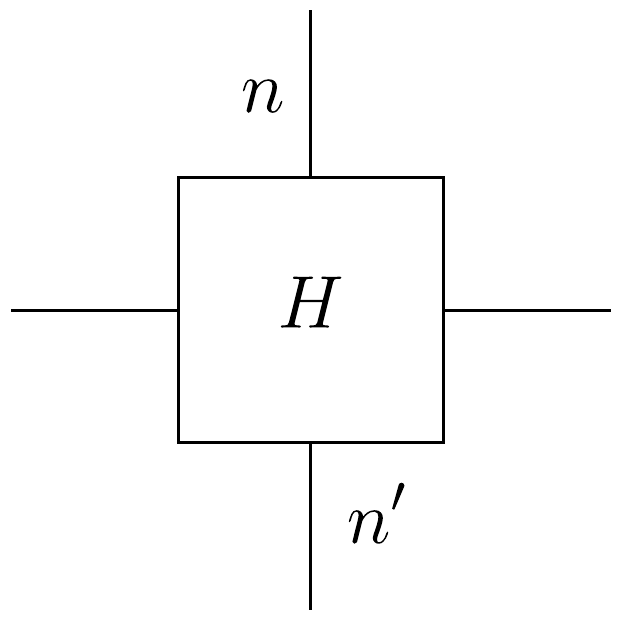}}}~~= \\
    \begin{pmatrix} 
    \mathbb{I} & -t e^{i\varphi}{\bf A}^\dagger & -te^{-i\varphi} {\bf A} & -(\mu+\frac{U}{2}){\bf N}+\frac{U}{2}{\bf N}^2 \\
    0 & 0 & 0 & {\bf A} \\
    0 & 0 & 0 & {\bf A}^\dagger\\
    0 & 0 & 0 & \mathbb{I}
    \end{pmatrix}
    \label{eq:mpo}
\end{multline}
where the operators $\mathbb{I}$, ${\bf A}^\dagger$, ${\bf A}$, ${\bf N}$ and ${\bf N}^2$ are represented as matrices in the number occupation basis $(n,n')$. The rows and columns of the right-hand-side of Eq.~(\ref{eq:mpo}) correspond to the left and right legs of the tensor ${\bf H}$, respectively. The Peierls phase, $\varphi$, arises from the gauge transformation discussed in Sec.~\ref{sec:calcSF} and is used to compute the superfluid density.  For the state to be normalized we require that $||{\bf \tilde W}||_2=||{\bf \tilde S}||_2=1$.  The square of this norm, which is basis independent, equals the sum of the modulus squared of all matrix elements.

The energies ${\cal E}_W$ and ${\cal E}_S$ are extensive, and hence formally infinite. As explained in Appendices~C and~D of Ref.~\cite{vumps}, these divergences can be subtracted off. We summarize the procedure for doing so below. The optimal $\bf \tilde W$ and $\bf \tilde S$ solve eigenvalue problems
%As the ${\bf L}$ and ${\bf R}$ tensors are known, it is a standard calculation~\cite{SCHOLLWOCKreview} to contract all tensors to the left and right of ${\bf \tilde W}$ in Eq.~(\ref{eq:make_heff}). The only wrinkle is the fact that there are an infinite number of terms, so $\langle \Psi_{\tilde W}|\hat{H}|\Psi_{\tilde W}\rangle$ is formally infinite. This is merely a consequence of the extensiveness of energy: in the limit of a large system size, the total energy of an $N$-site chain is $E_N\approx N\varepsilon$ (neglecting boundary effects) where $\varepsilon$ is the energy per site in the thermodynamic limit. Intuitively, one can remove this divergence by subtracting $\varepsilon$ from each tensor ${\bf H}$. This is precisely what we do, although we follow the more efficient procedure laid out in Appendices~C and~D of Ref.~\cite{vumps}.
%
%We now choose ${\bf \tilde W}$ such that Eq.~(\ref{eq:make_heff}) is minimized, subject to the constraint that the MPS is normalized (i.e. $||{\bf \tilde W}||_2=1$). This is equivalent to choosing ${\bf \tilde W}$ to be the lowest-energy eigenvector of the effective Hamiltonian,
\begin{align}
    \vcenter{\hbox{\includegraphics[scale = 0.3]{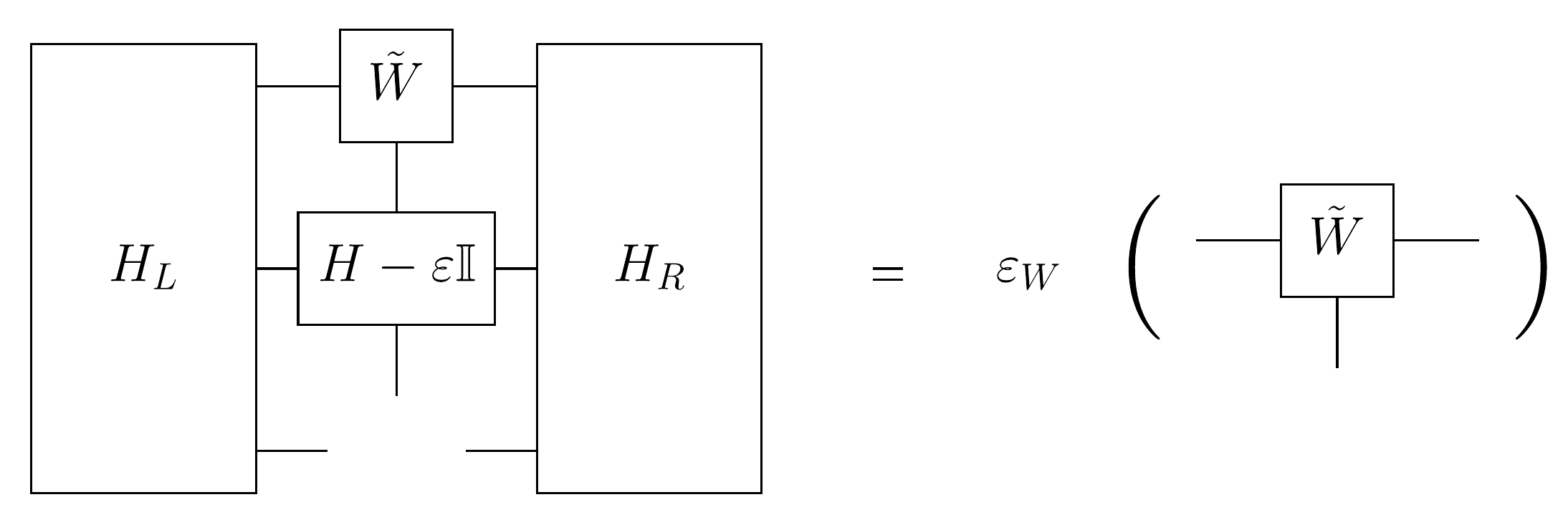}}},
    \label{eq:eigW}\\
    \vcenter{\hbox{\includegraphics[scale = 0.32]{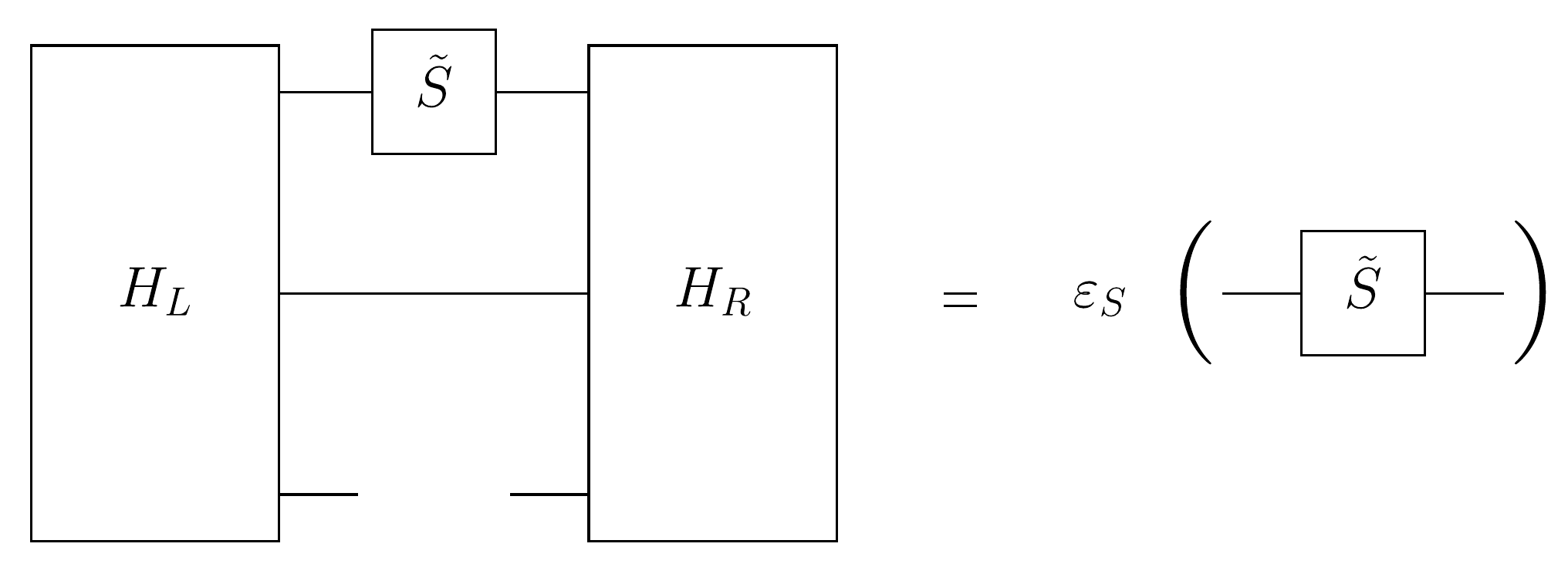}}},
    \label{eq:eigS}
\end{align}
where ${\bf H}-\varepsilon \mathbb{I}$ is the Hamiltonian MPO minus the average energy per site, $\varepsilon$ (which is defined in Eq.~(\ref{eq:vumpsEnergy})). This simply requires modifying the on-site term of Eq.~(\ref{eq:mpo}) to be $-(\mu+U/2){\bf N}+(U/2){\bf N^2}-\varepsilon \mathbb{I}$.
The tensors ${\bf H_L}$ and ${\bf H_R}$ consist of all contributions to the left and right %(respectively) 
of the central tensor in Eq.~(\ref{eq:make_heff}), with the same subtraction %after subtracting off the extensive contribution 
\cite{vumps}. 
%{\color{blue} The tensor ${\bf H}-\varepsilon \mathbb{I}$ is the Hamiltonian MPO minus the average energy per site, $\varepsilon$ (see Eq.~(\ref{eq:vumpsEnergy})). This is obtained by modifying the on-site term of Eq.~(\ref{eq:mpo}) to be $-(\mu+U/2){\bf N}+(U/2){\bf N^2}-\varepsilon \mathbb{I}$.}
For convenience, we will refer to the eigenvalue problems in Eq.~(\ref{eq:eigW}) and (\ref{eq:eigS}) as ${\bf H_W}({\bf \tilde W})=\varepsilon_W{\bf \tilde W}$ and ${\bf H_S}({\bf \tilde S})=\varepsilon_S{\bf \tilde S}$.  
%The divergence subtraction procedure forces ${\bf H_W}({\bf W})={\bf 0}$ and ${\bf H_S}({\bf S})={\bf 0}$.

The tensors ${\bf H_L}$ and ${\bf H_R}$ can be immediately evaluated using Eqs.~(\ref{eq:make_heff}) and~(\ref{eq:mpo}). Just as ${\bf H}$ was written as a matrix of matrices in Eq.~(\ref{eq:mpo}), ${\bf H_L}$ and ${\bf H_R}$ can be represented as vectors of matrices. For example, %before subtracting off the divergence, 
${\bf H_L}$ is given by
\begin{equation}
    \begin{pmatrix}
    \vcenter{\hbox{\includegraphics[scale = 0.2]{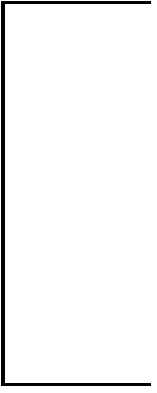}}}~~ &
    -te^{i\varphi}\vcenter{\hbox{\includegraphics[scale = 0.2]{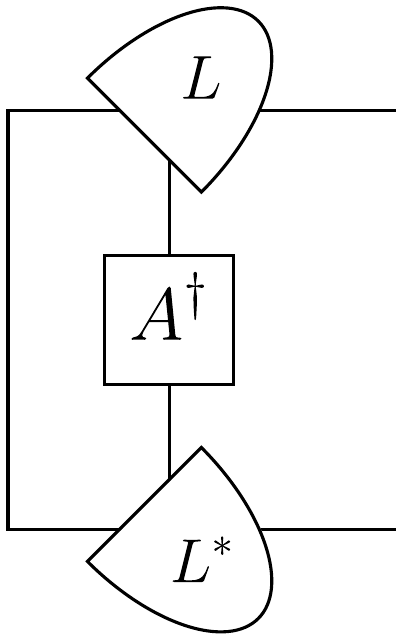}}}~~ &
    -te^{-i\varphi}\vcenter{\hbox{\includegraphics[scale = 0.2]{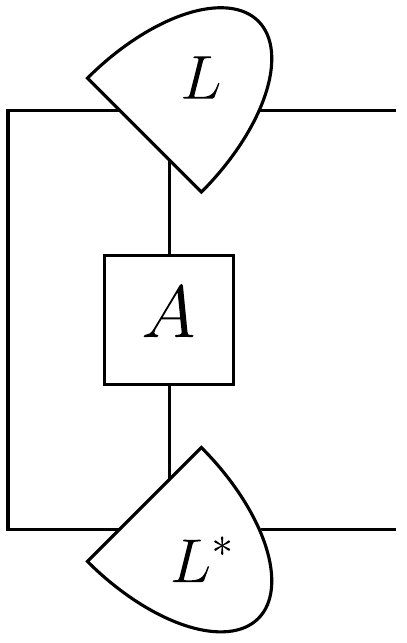}}}~~~~~ &
    \vcenter{\hbox{\includegraphics[scale = 0.2]{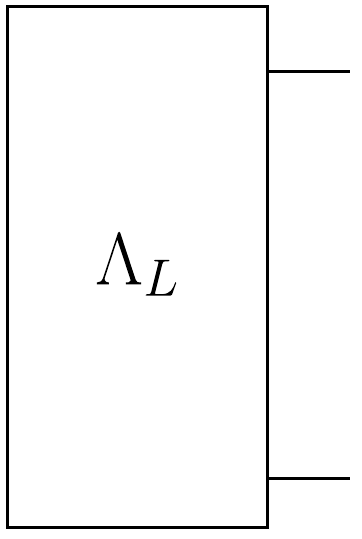}}}
    \end{pmatrix}
    \label{eq:HL}
\end{equation}
where ${\bf \Lambda_L}$ encompasses all terms in the Hamiltonian that act on sites to the left of the central site. Before subtracting off the divergences, these ``disconnected" terms are given by
\begin{equation}
    \begin{aligned}
    \vcenter{\hbox{\includegraphics[scale = 0.33]{LambdaL.pdf}}} ~ = ~& -te^{i\varphi}\vcenter{\hbox{\includegraphics[scale = 0.33]{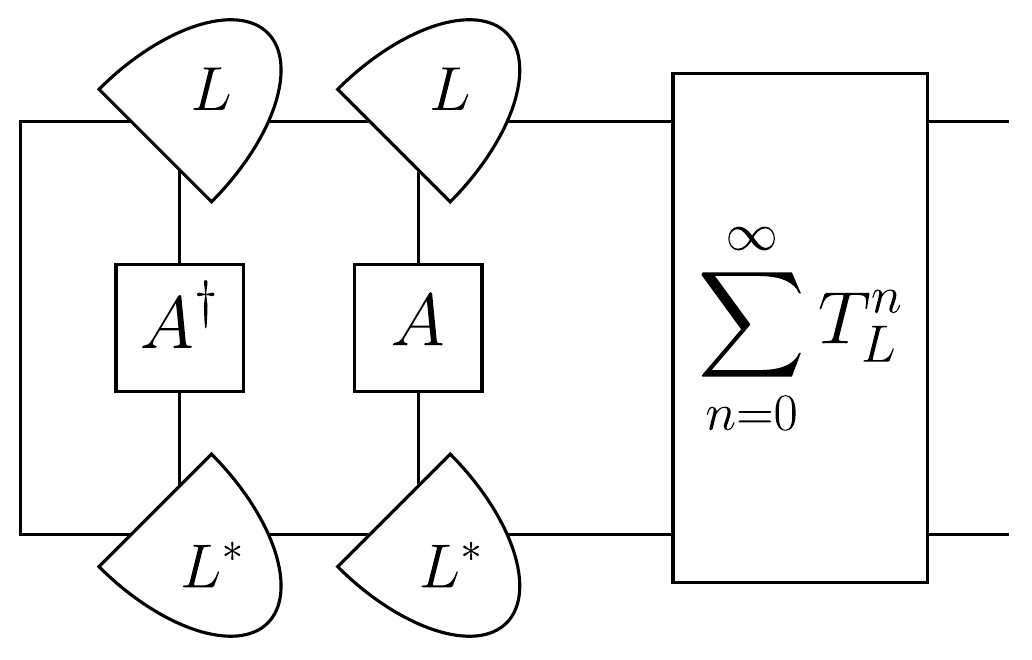}}}+h.c.\\
    &-(\mu+U/2)~\vcenter{\hbox{\includegraphics[scale = 0.33]{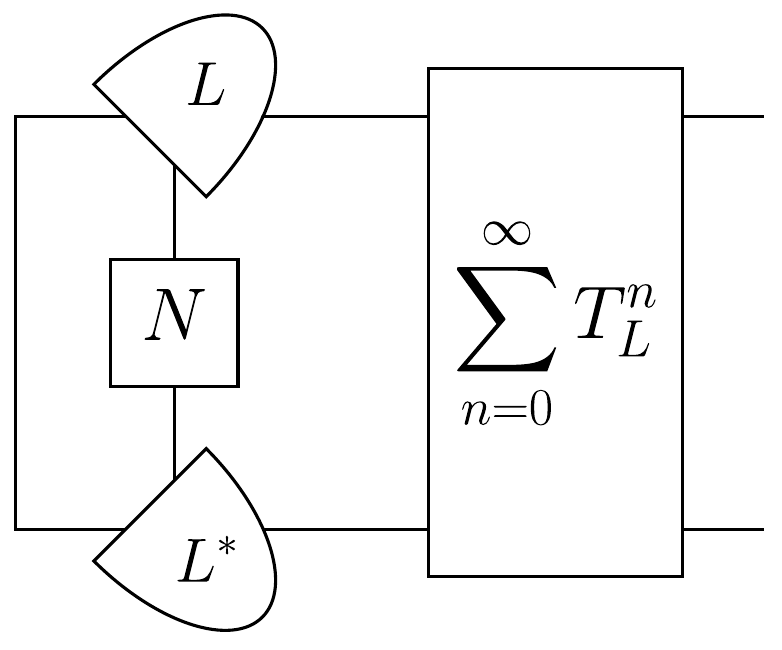}}} \\
    &~+(U/2)~\vcenter{\hbox{\includegraphics[scale = 0.33]{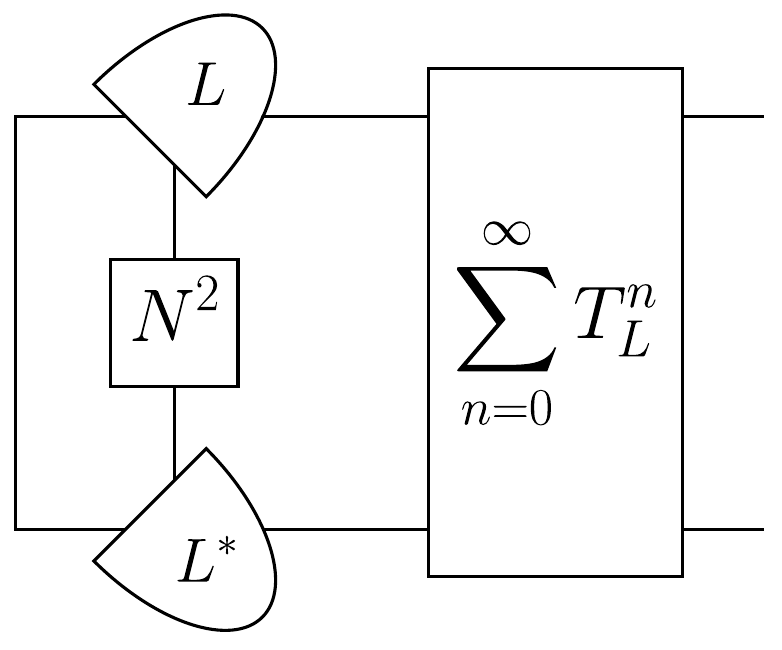}}} \\
    \equiv ~& \vcenter{\hbox{\includegraphics[scale = 0.33]{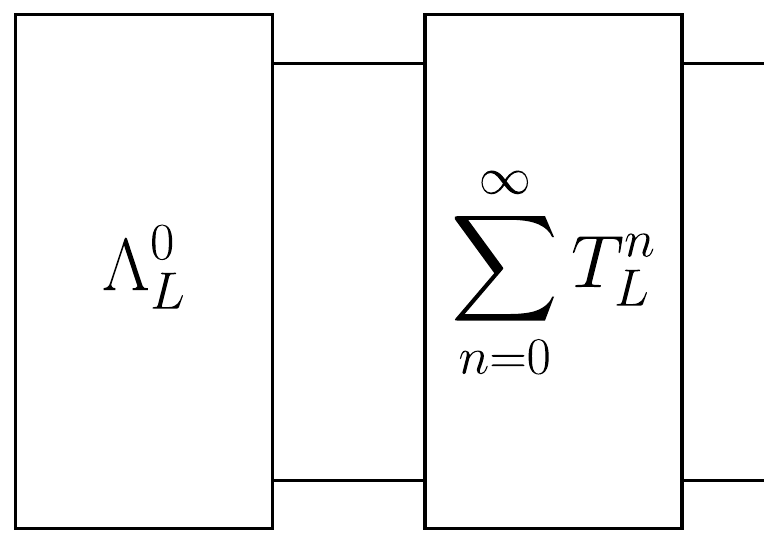}}}.
    \end{aligned}
    \label{eq:LamL}
\end{equation}
which involves the geometric sum 
$\sum_{n=0}^\infty {\bf T_L}^n=({\bf 1}-{\bf T_L})^{-1}$ 
where ${\bf T_L}$ is the left-canonical transfer matrix:
\begin{equation}
    \vcenter{\hbox{\includegraphics[scale = 0.45]{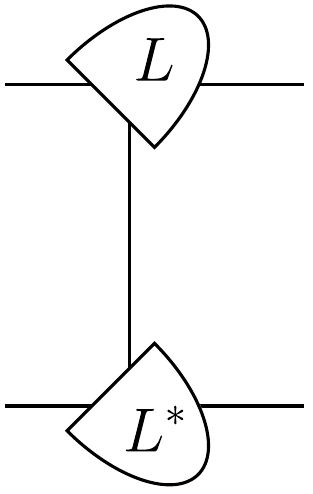}}} ~~=~~ \vcenter{\hbox{\includegraphics[scale = 0.5]{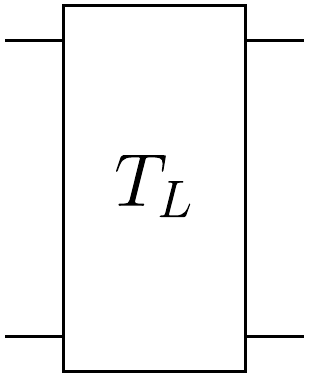}}}.
\end{equation}
Subtracting off the divergence formally requires replacing
\begin{equation}
    \vcenter{\hbox{\includegraphics[scale = 0.3]{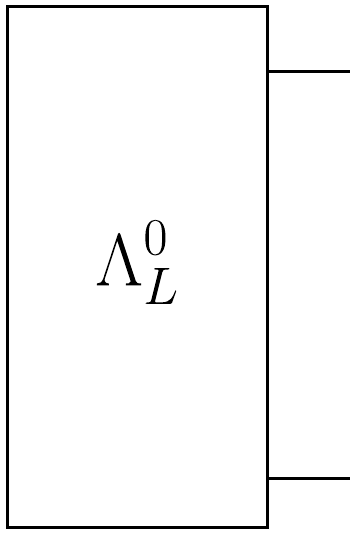}}} ~~\to~~ \vcenter{\hbox{\includegraphics[scale = 0.3]{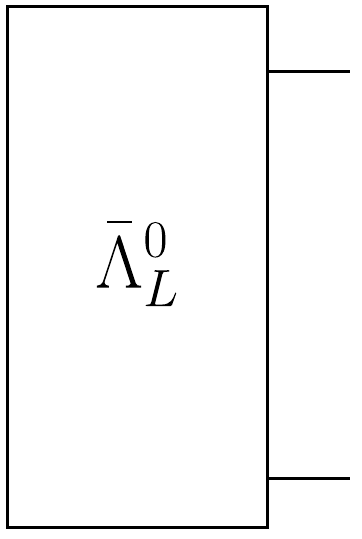}}} ~~=~~
    \vcenter{\hbox{\includegraphics[scale = 0.3]{LambdaL0.pdf}}} ~~-~~
    \varepsilon~
    \vcenter{\hbox{\includegraphics[scale = 0.32]{id.pdf}}}.
\end{equation}
In Eq.~(\ref{eq:LamL}) the divergence can be associated with the fact that the transfer matrix has an eigenvector with eigenvalue 1.  This suggests an alternative renormalization, substituting ${\bf T_L}\to{\bf \bar T_L}$ with ${\bf \bar T_L}={\bf T_L}-|0_L)(0_L|$, where $(0_L|$ and $|0_L)$ are the dominant left and right eigenvectors of ${\bf T_L}$. When Eq. (\ref{eq:ti_equality}) is satisfied, these are given by
\begin{align}
    (0_L| &~=~ \vcenter{\hbox{\includegraphics[scale = 0.45]{id.pdf}}} \hspace{1.5cm}
    |0_L) ~=~ \vcenter{\hbox{\includegraphics[scale = 0.33]{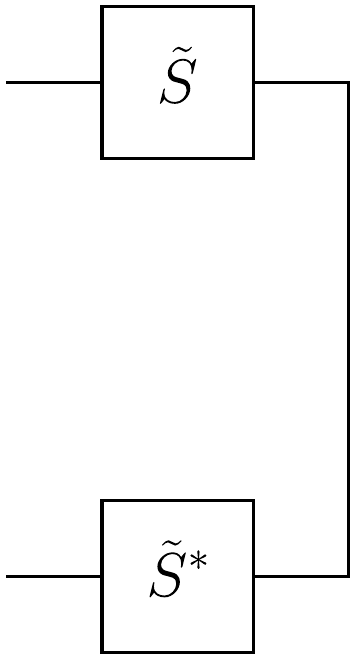}}}.
\end{align}

To show the equivalence of these approaches, we note that the average energy per site is
\begin{equation}
    \varepsilon = \Lambda_L^0|0_L) ~ = ~ \vcenter{\hbox{\includegraphics[scale = 0.33]{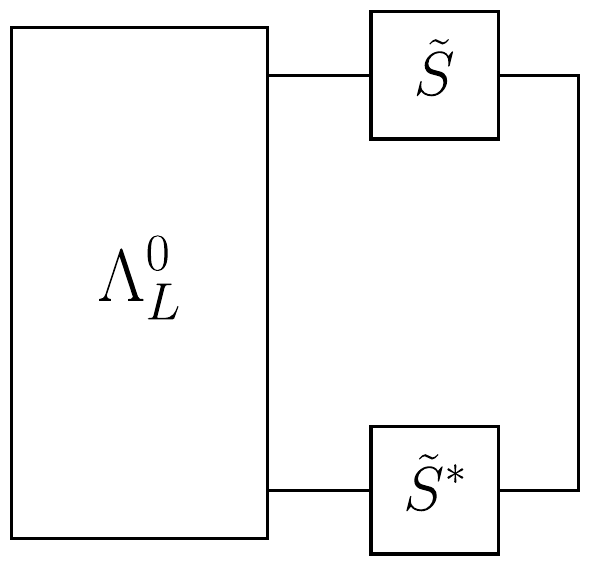}}}.
    \label{eq:vumpsEnergy}
\end{equation}
Hence the identity ${\bf \bar \Lambda_L^0} {\bf T_L}^n={\bf \Lambda_L^0} {\bf \bar T_L}^n$ can be applied to each term in the geometric sum for $n\geq1$ (see Eq.~(\ref{eq:LamL})). This construction implies that ${\bf H_L}$ and ${\bf H_R}$ are the fixed points of the left and right MPO tranfer matrices, respectively:
%, defined graphically below:
\begin{align}
    \vcenter{\hbox{\includegraphics[scale = 0.33]{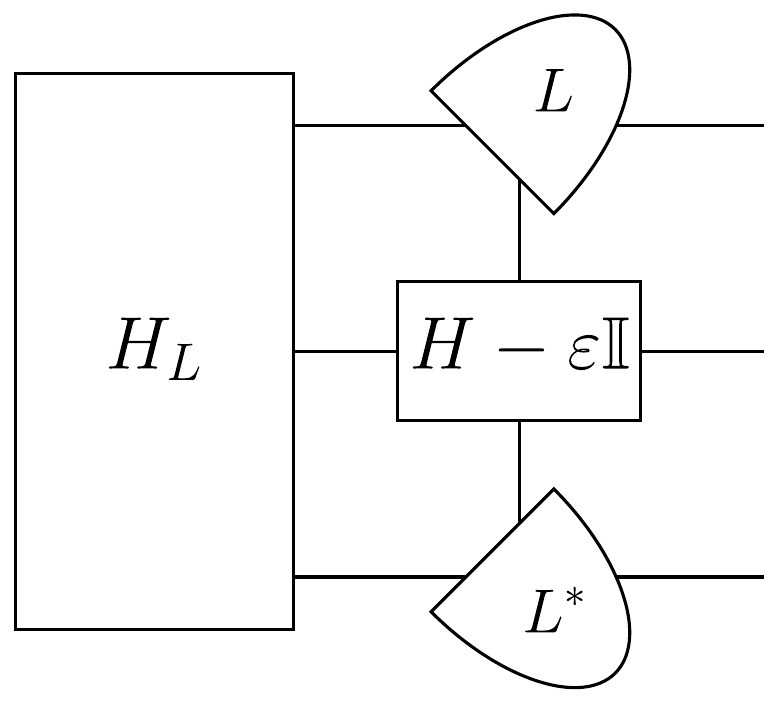}}} ~~=~~ \vcenter{\hbox{\includegraphics[scale = 0.32]{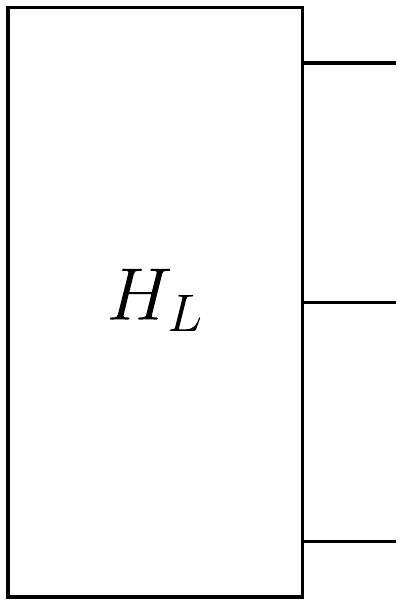}}} \\
    \vcenter{\hbox{\includegraphics[scale = 0.33]{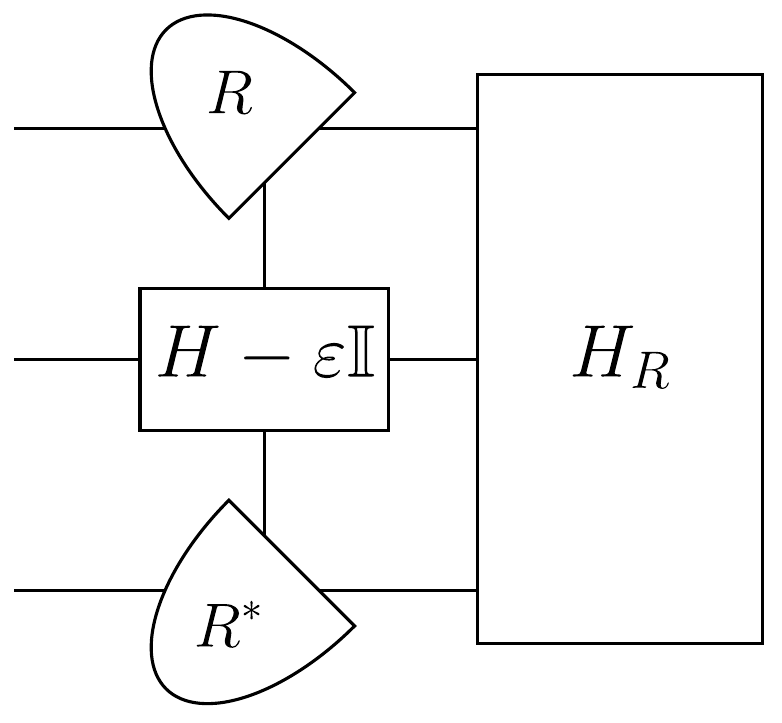}}} ~~=~~ \vcenter{\hbox{\includegraphics[scale = 0.32]{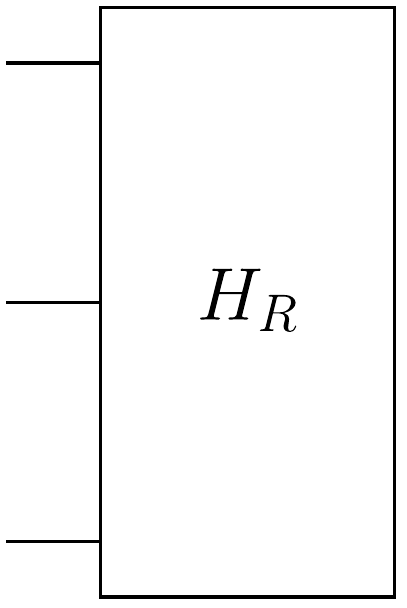}}}.
\end{align}
We then find ${\bf \tilde L}$ and ${\bf \tilde R}$ by minimizing
\begin{align}
    \vcenter{\hbox{\includegraphics[scale = 0.33]{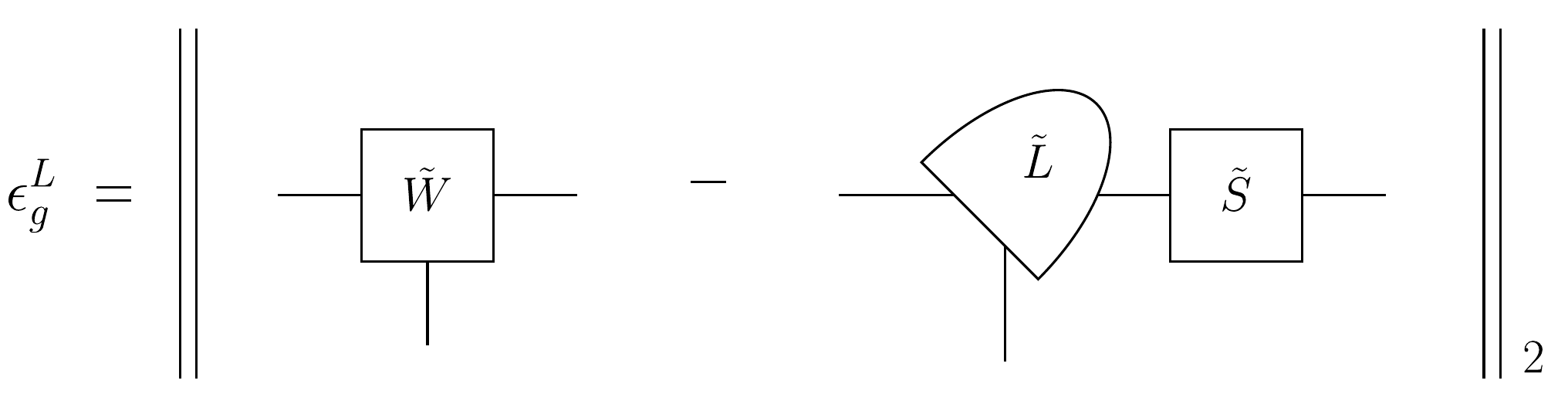}}}
    \label{eq:ti_l}\\
    \vcenter{\hbox{\includegraphics[scale = 0.33]{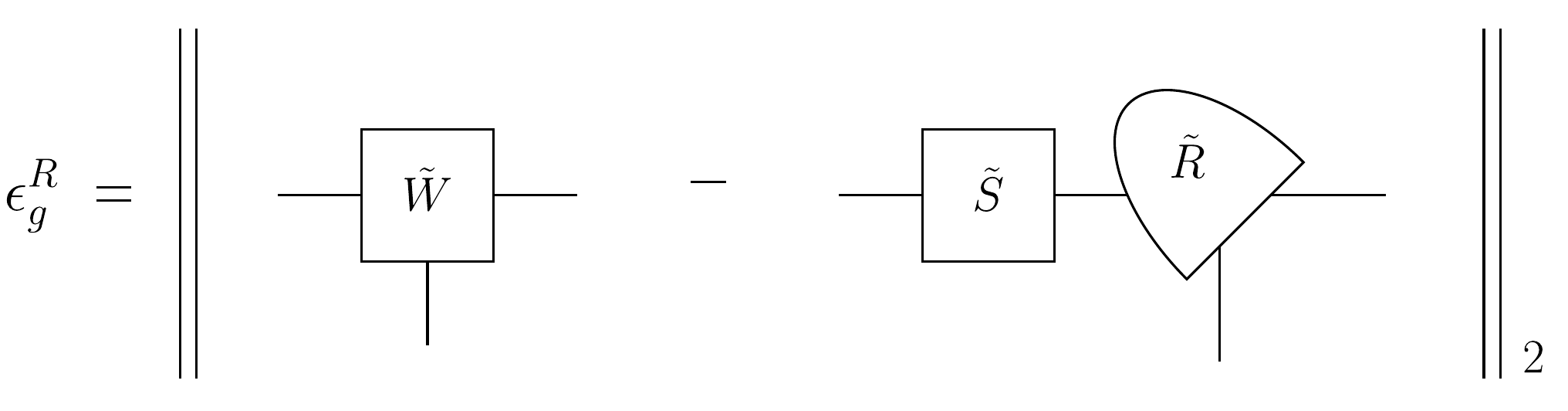}}}
    \label{eq:ti_r}.
\end{align}
The gauge-fixing error, defined as $\epsilon_g={\rm max}\{ \epsilon_g^L, \epsilon_g^R \}$, quantifies the deviation of the mixed-canonical MPS (defined by tensors ${\bf \tilde L}$, ${\bf \tilde S}$, and ${\bf \tilde R}$) from translational invariance. As the VUMPS algorithm is iterated, we find that $\epsilon_g$ decreases and eventually approaches machine precision.

The optimization in Eq.~(\ref{eq:ti_l}) and (\ref{eq:ti_r}) can be performed exactly using two
%where the tensors ${\bf \tilde L}$ and ${\bf \tilde R}$ are chosen such that $\epsilon_g^L$ and $\epsilon_g^R$ are minimized. The solution to this optimization problem is known and can be found using two 
singular-value decompositions. We refer the reader to Sec.~II~C of Ref.~\cite{vumps} for the expression and for an approximation that better handles singular values near machine precision.

%exact solution. Given the tensors ${\bf \tilde L}$, ${\bf \tilde S}$ and ${\bf \tilde R}$, the (approximately) uniform MPS that forms the basis for the next iteration is defined.

%The gauge-fixing error, defined as $\epsilon_g={\rm max}\{ \epsilon_g^L, \epsilon_g^R \}$, quantifies the deviation of the mixed-canonical MPS (defined by tensors ${\bf \tilde L}$, ${\bf \tilde S}$, and ${\bf \tilde R}$) from translational invariance. As the VUMPS algorithm is iterated, we find that $\epsilon_g$ vanishes and eventually approaches machine precision.

%In Refs.~\cite{vumps,vumps2}, the authors 

If $\epsilon_g=0$, the distance from the optimal variational ansatz can be quantified by calculating the magnitude of the gradient of the energy with respect to ${\bf W}$, constrained to the manifold of uniform states. As argued in Refs.~\cite{vumps,vumps2}, this gradient can be expressed as
%In addition to the gauge-fixing error, one might want to quantify how well the mixed-canonical state approximates the ground state of the Hamiltonian. In general, the energy of a variational ansatz $|\Psi_{\rm var}(\{ a_i \})\rangle$ with energy $E(|\Psi\rangle)=\langle \Psi|\mathcal{H}|\Psi\rangle/\langle \Psi|\Psi\rangle$ is a function of the set of variational parameters, $\{ a_i \}$. The variational ground state is the state whose energy is the absolute minimum of $E_{\rm var}(\{a_i\})$ with respect to those parameters. If the variational states comprise a manifold, as is the case for the set of uniform MPS~\cite{vumps}, then the energy of the variational ground state is defined has having a vanishing parameter-space gradient: $|\nabla_{a_i}E_{\rm var}(\{a_i\})|=0$. Hence, we can use the norm of the gradient as a measure of how close a given state is to the variational minimum. Refs.~\cite{vumps,vumps2} compute the gradient, ${\bf G}$, explicitly using tangent space techniques, finding that 
\begin{equation}
    \vcenter{\hbox{\includegraphics[scale = 0.25]{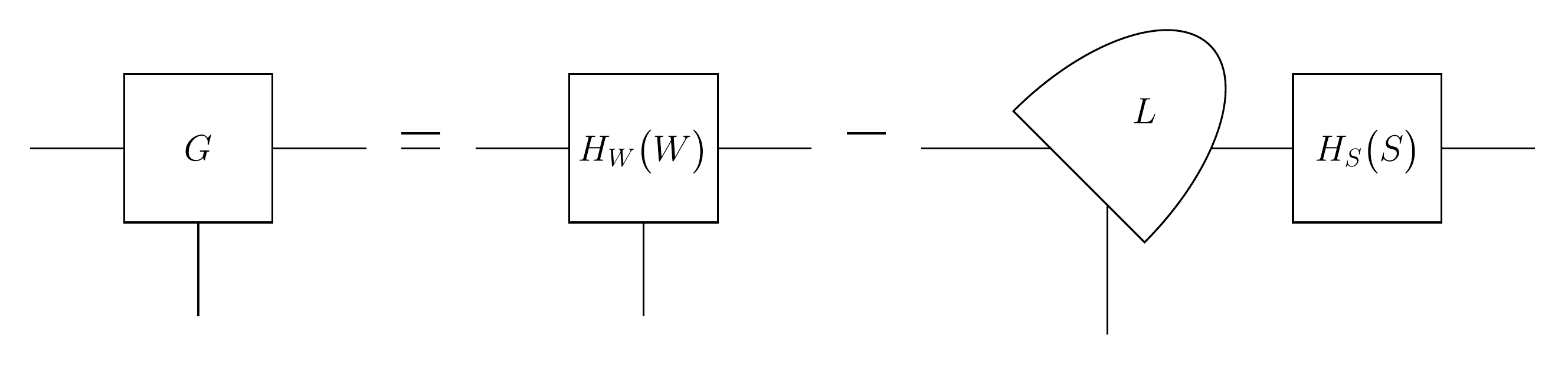}}},
    \label{eq:gradient}
\end{equation}
where we defined the tensors ${\bf H_{W}}$ and ${\bf H_S}$ after Eqs.~(\ref{eq:eigW}) and~(\ref{eq:eigS}). 
%Note that the gradient of the state defined by tensors $\{ {\bf L}, {\bf S}, {\bf R}, {\bf W} \}$ involves ${\bf H_W(W)}$ rather than ${\bf H_W(\tilde W})$, as in Eq.~(\ref{eq:eigW}). 
%, respectively, although there is an ambiguity due to the fact that $E_W$ and $E_S$ have not been defined. For the purpose of Eq.~(\ref{eq:gradient}), we define the effective Hamiltonian tensors such that ${\bf H_W}({\bf W})={\bf 0}$ and ${\bf H_S}({\bf S})={\bf 0}$, i.e. we subtract the energy per site off of each MPO operator ${\bf H}$ in the construction of ${\bf H_W}$ and ${\bf H_S}$. This implies that $E_W$ and $E_S$ vanish as the state converges.
The norm of the gradient, $g\equiv||{\bf G}||_2$, vanishes at the variational minimum.
Even when $\epsilon_g\neq0$, the quantity $g$ has meaning, and we quantify our proximity to the optimal state by the magnitude of $g$.
%We compute the norm of the gradient, $g\equiv||{\bf G}||_2$, after each iteration in order to quantify how close our state is to the variational minimum.
%Even when $\epsilon_g\neq0$, the quantity $g$ has meaning.
%Note that $g$ formally vanishes in the limit that the energy converges and $\epsilon_g$ vanishes.
In practice, the variational energy converges to within machine precision much faster than $g$. For the purpose of this paper, we define convergence as $g\leq 10^{-14}$.

One of the strengths of VUMPS is it can make large steps in parameter-space.  Unfortunately, the algorithm sometimes stalls out or falls into a limit cycle.
%, with $g$ saturating at
%Interestingly, we found that the VUMPS algorithm would sometimes struggle to reduce $g$ below some threshold, which was often of the order 
%$g\sim 10^{-11}$. 
When this was the case, we were able to reduce $g$ to the desired precision by applying state updates using the infinite time-dependent variational principle (iTDVP)~\cite{vumps2}. The procedure is very similar to VUMPS except that, instead of solving for the lowest-energy eigenvector of ${\bf H_{W}}$, we update the state by defining ${\bf \tilde W}=e^{-\tau {\bf H_W}}({\bf W})$ where $\tau$ is an imaginary time step and ${\bf H_W}$ is defined in Eq.~(\ref{eq:eigW}). Of course, the normalization condition ($||{\bf \tilde W}||_2=1$) must now be enforced by hand. Similarly, we update ${\bf \tilde S}=e^{-\tau{\bf H_S}}({\bf S})$. In the limit $\tau\to\infty$, iTDVP state updates and VUMPS state updates are equivalent. One can then proceed as we did with VUMPS, defining ${\bf \tilde L}$ and ${\bf \tilde R}$ according to Eqs.~(\ref{eq:ti_l}) and~(\ref{eq:ti_r}) and computing the gradient using Eq.~(\ref{eq:gradient}). 

The iTDVP algorithm should reliably converge to the ground state for small $\tau$, although small time steps also mean that more iterations will be required to reach the variational ground state. We deployed iTDVP updates in two ways: (1) when VUMPS updates would not take $g$ below some threshold, most often $g\sim 10^{-11}$, iTDVP updates with $\tau\sim \mathcal{O}(1)$ could reduce $g$ below our convergence criterion; and (2) when the algorithm was prone to falling into limit cycles, we used iTDVP updates with $\tau\sim \mathcal{O}(0.1)$ in between successive VUMPS updates to improve convergence.

\section{\label{sec:quasicondensate}Quasicondensate density}
While the 1D Bose-Hubbard model has zero condensate density, a consequence of the Mermin-Wagner theorem~\cite{merminWagner,hohenbergLRO}, simulations of the model in finite-sized systems will observe a finite quasicondensate density, $\rho_{qc}=|\langle a_i\rangle|^2$. Our simulations are performed in the thermodynamic limit but make use of variational wavefunctions with finite correlation lengths, producing an analogous effect. One can deduce the asymptotic bond-dimension scaling of the quasicondensate density by approximating the density matrix $\langle a_ia^\dagger_j\rangle$ as a piecewise function that decays as $|i-j|^{-K/2}$ for $|i-j|\leq\xi$ and is constant for $|i-j|>\xi$. Making use of Eq.~(\ref{eq:kappa}), this cartoon yields a quasicondensate density that scales as
\begin{equation}\label{qcscale}
    \rho_{qc}\propto \langle a_0a^\dagger_\xi\rangle\propto\chi^{-\kappa K/2}.
\end{equation}

\begin{figure}
    \centering
    \includegraphics[width=3.375in]{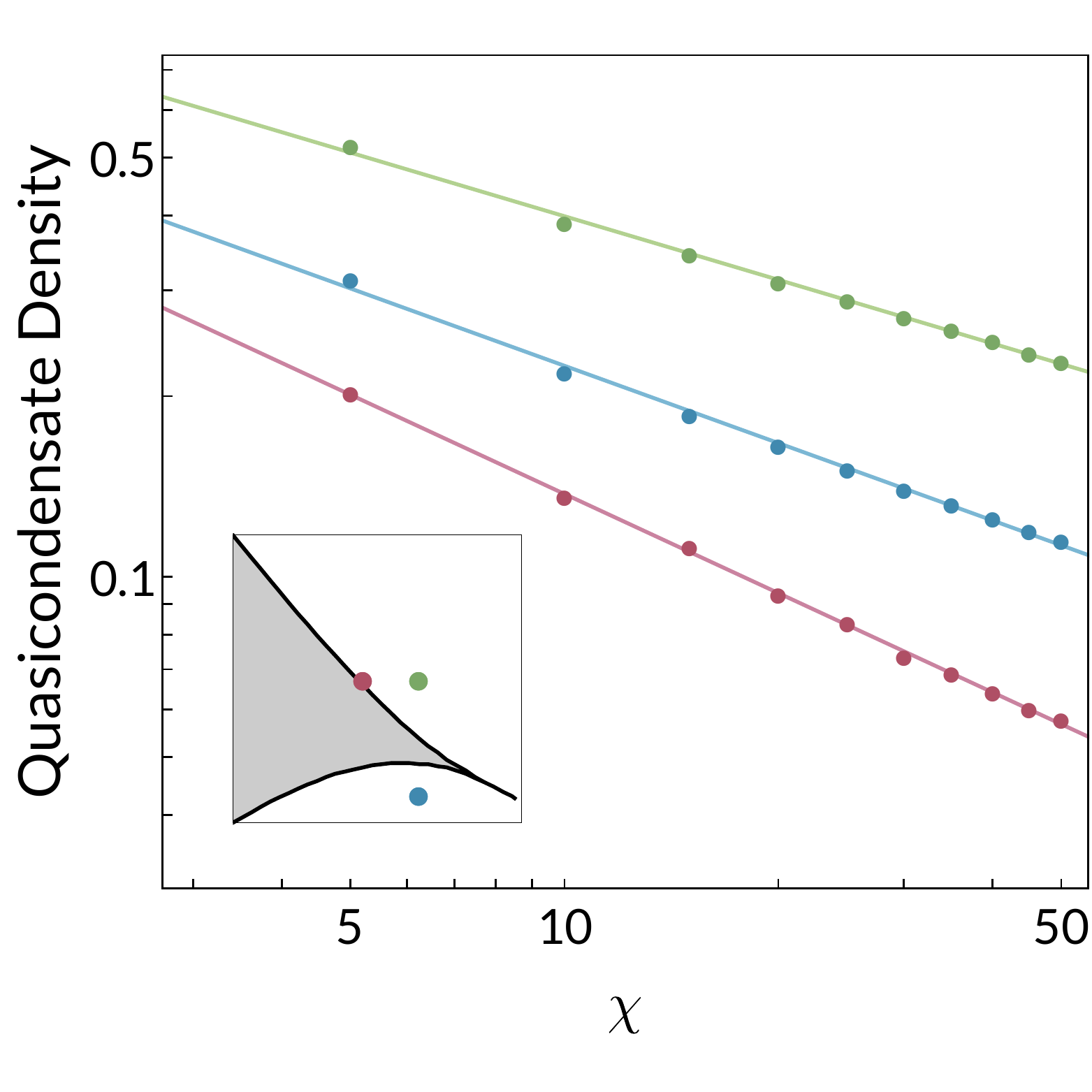}
    \caption{Quasicondensate density, $\rho_{qc}=|\langle a_i\rangle|^2$, versus bond dimension on a log-log scale for a few representative points in the Luttinger liquid phase. Solid lines give fits of the form $\rho_{qc}=\alpha\chi^{-\kappa K/2}$ where $\alpha$ is the only free parameter.}
    \label{fig:quasiCondFig}
\end{figure}

In Figure~\ref{fig:quasiCondFig} we plot $\rho_{qc}$ versus bond dimension on a log-log scale. The quasicondensate density decays as a power law, as expected. The solid lines give fits to the data of the form $\rho_{qc}(\chi)=\alpha\chi^{-\kappa K/2}$ where $\alpha$ is the only free parameter ($K$ is determined from $\langle a_ia^\dagger_j\rangle$, see Appendix~\ref{sec:paramFits}). 
The quality of the fits are strong confirmation of Eq.~(\ref{qcscale}).
%In particular, the 
%power law exponent is fixed by our prior calculation of $K$.
%captured extremely well by this scaling argument.
%made above. %As noted at the end of Sec.~\ref{sec:entanglement}, this is an interesting example in which the scaling of a quantity with bond dimension depends both on the Luttinger parameter, $K$, and the conformal charge, $c$ (vis-\`a-vis $\kappa$, see Eq.~(\ref{eq:kappa})).

\section{\label{sec:paramFits}Determining the Luttinger parameter}
The Luttinger parameter, $K$, characterizes many of the properties of a Luttinger liquid. As such, there are a variety of ways to determine the Luttinger parameter of a uniform MPS. In Figure~\ref{fig:luttingerParam} we plot $K$, computed in three different ways, as a function of $\mu/U$. In this Appendix we compare these methods and discuss their reliability. 

Data is taken at fixed $t/U=0.15$ and using the converged uMPS at bond dimensions $\chi=20$, 30 and 40. The vertical line at %dashed cross section approaches the Mott-SF transition at 
$\mu/U\approx 0.445$, denotes the Mott transition, where $K\to 1$~\cite{fisherBH,giamarchi}.

\begin{figure}
    \centering
    \includegraphics[width=3.375in]{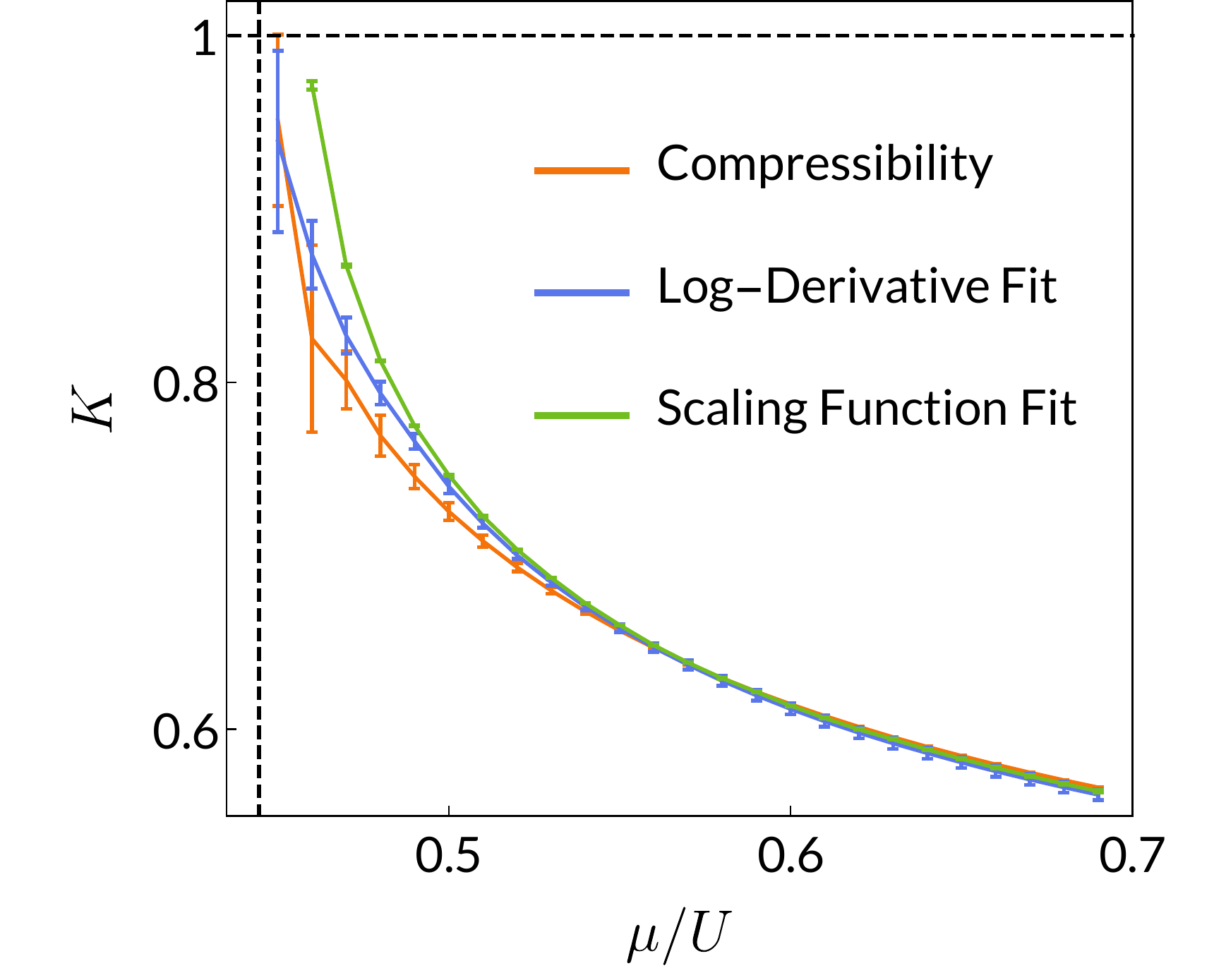}
    \caption{Plot of the Luttinger parameter versus $\mu/U$ at fixed $t/U=0.15$, determined using three different procedures. The Mott-SF transition point is at $\mu/U\approx 0.445$, denoted by the black dashed line. See Appendix~\ref{sec:paramFits} for an explanation of the procedures.}
    \label{fig:luttingerParamFits}
\end{figure}

The orange curve is determined from $K=\sqrt{v_n/v_j}$, where $v_n$ and $v_j$ are defined in Sec.~\ref{sec:LL}.  We compute $v_j$ from superfluid density, calculated using the approach in Sec.~\ref{sec:calcSF}, and the relationship
\begin{equation}
    2t\rho_s=\frac{\hbar v_j}{\pi}.
\end{equation}
We compute $v_n$ from %the 
%computed at each point, as well as the compressibility. We approximate 
the compressibility, $\kappa=\partial n/\partial\mu$, calculated using finite differences.
We measure the density as a function of chemical potential and then apply
\begin{equation}
\kappa=\frac{1}{\hbar\pi v_n}.
\end{equation}
%by measuring the density while we sweep the chemical potential. 
Note that this method was also employed in Fig.~\ref{fig:luttingerParam}. 
%As discussed in the main text, $\rho_s$ and $\kappa$ are related to the characteristic velocities of phase and density fluctuations, respectively, via
%\begin{equation}
%    2t\rho_s=\frac{\hbar v_j}{\pi}\hspace{1.5cm}\kappa=\frac{1}{\hbar\pi v_n}.
%\end{equation}
%The Luttinger parameter is then given by $K=\sqrt{v_n/v_j}$. 
We establish error bars on the superfluid density by measuring the variance of $\rho_s(\chi)$ at the three different bond dimensions. In general, however, the error bars on the orange curve are dominated by errors in the discrete derivative used to calculate the compressibility.

The blue and green curves in Fig.~\ref{fig:luttingerParamFits} are both determined from the algebraic decay of the density matrix, $\langle a_ia^\dagger_j\rangle$, plotted in Fig.~\ref{fig:densityMatFig}. In the blue curve, we take the derivative of $\ln\langle a_ia^\dagger_j\rangle$ with respect to $\ln|i-j|$ and find the average value where the curve plateaus. We establish error bars by taking the standard deviation of the log-derivative over the domain $\ln(|i-j|/\xi)\in(-2,0)$, where $\xi(\chi)$ is determined by Eq.~(\ref{eq:xi}). Note that the power-law behavior breaks down for $|i-j|>\xi$, beyond which $\langle a_ia^\dagger_j\rangle$ decays exponentially to $\rho_{qc}$.

The green curve is determined by rescaling $\langle a_ia^\dagger_j\rangle$ by a power of the correlation length such that the $\chi=20$, 30 and 40 curves exhibit a scaling collapse. The collapsed curves are then fit to a scaling function of the form~\cite{sethnaScalingFunc} 
\begin{equation}
    C(x)=a\big(1 + (x/b)^{-n\eta}\big)^{1/n}
    \label{eq:Cx}
\end{equation}
using a non-linear least squares algorithm. In practice, in order to arrive at an unbiased scaling collapse, we exploit the fact that the collapse should occur when we rescale the axes as follows: $|i-j|\to|i-j|/\xi$ and $\langle a_ia^\dagger_j\rangle\to\langle a_ia^\dagger_j\rangle\xi^{K/2}$. Furthermore, the parameter $\eta$ in Eq.~(\ref{eq:Cx}) should be equal to $K/2$ at convergence. We therefore implement an iterative scheme to find the optimal value of $K$: we start by rescaling the curves by an arbitrary power of $\xi$; we then fit the data to a scaling function and extract the Luttinger parameter $\tilde K=2\eta$; we then use $\tilde K$ to rescale the curves and repeat the process. We need about 5 iterations to reach convergence. Error bars come from the covariance matrix of the non-linear least squares fit, which we then rescale to account for systematic errors in the fitting procedure~\cite{sethnaUncertainty}.

We find that the orange and blue curves agree reasonably well within their error bars for all data points. Errors in the orange curve increase near the Mott-SF transition because the curvature of $n(\mu)$ increases, making the discrete derivative approximation less accurate. The error could be improved substantially by taking data at more finely-spaced values of $\mu$. As for the blue and green curves, the density matrix develops oscillations that persist to longer and longer distances as one approaches the phase boundary. 
%As we approach the Mott-SF boundary, t
When this length-scale exceeds the correlation length of the uMPS  it becomes challenging to extract $K$ from 
$\langle a_ia^\dagger_j\rangle$.
%this behavior overwhelms the algebraic decay out to distances beyond the correlation length of the uMPS, the relationship between $\langle a_ia^\dagger_j\rangle$ and $K$ can no longer be applied. 
In this case, the log-derivative technique (blue) yields large error bars that likely encompass the correct value of $K$. The green curve, on the other hand, systematically overfits based on this behavior and deviates significantly from the other two curves.  A second consequence of the overfitting is that the error bars on the green curve become unreliable near the transition.
%Note as well that error bars on the green curve are unrealistically small near the transition, a consequence of overfitting. 
The accuracy of both the blue and green curves would be substantially improved by working at larger bond dimensions, where the correlation length is larger.
%predicted power-law behavior (Eq.~(\ref{eq:dMatLL})) could be observed before $|i-j|\sim \xi$.

% \bibliographystyle{apsrev4-2}
% \bibliography{bib.bib}


\end{document}